\documentclass[
    aps,
    pra,
    longbibliography,
    superscriptaddress,amsmath,amssymb,amsfonts, 
    tightenlines,
    twocolumn,
    ]{revtex4-2}

\usepackage[utf8]{inputenc}  
\usepackage[T1]{fontenc}     
\usepackage[english]{babel}  

\usepackage[colorlinks = true, linkcolor=teal, citecolor=teal, urlcolor  = teal]{hyperref}

\usepackage{graphicx} 
\usepackage[babel]{microtype}  
\usepackage{amsmath,amssymb,amsthm,bm,amsfonts,mathrsfs,bbm} 
\usepackage{setspace}
\usepackage{times}
\usepackage{braket}
\usepackage[bottom]{footmisc}
\usepackage{blochsphere}
\usepackage{pgfplots}
\usepackage{amsthm}

\pgfplotsset{compat=1.17}
\usetikzlibrary{calc,3d,shapes, pgfplots.external, intersections}
\usepackage{algorithm}
\usepackage{physics}
\usepackage[noend]{algorithmic}
\usepackage{eqparbox}

\usepackage{tikz}
\usepackage{mathtools}
\usetikzlibrary{arrows.meta}
\usepackage{csquotes}

\usepackage{xspace}  
\usepackage{pgfplots}
\usepackage{verbatim}
\usepackage{amsmath}

\newtheorem{theorem}{Theorem}

\newtheoremstyle{note}
{3pt}
{3pt}
{}
{}
{\itshape}
{:}
{.5em}
{}

\newcommand\id{\leavevmode\hbox{\small1\kern-3.3pt\normalsize1}}

\newcommand{\x}{{\bf x}}

\definecolor{jens}{rgb}{0.1,0.4,0.6}

\usepackage{pdfpages}
\usepackage{pgffor}
\makeatletter
\AtBeginDocument{\let\LS@rot\@undefined}
\makeatother

\begin{document}

\title{Towards efficient quantum algorithms for diffusion probabilistic models}

\author{Yunfei Wang}
\email{Equal contribution}
\affiliation{Joint Center for Quantum Information and Computer Science, NIST/University of Maryland, College Park, MD 20742, USA}
\affiliation{Maryland Center for Fundamental Physics, Department of Physics, University of Maryland, College Park, MD 20742, USA}

\author{Ruoxi Jiang}
\email{Equal contribution}
\affiliation{Artificial Intelligence Innovation and Incubation Institute, Fudan
University, Shanghai 200437, China}

\author{Yingda Fan}
\affiliation{Department of Computer Science, The University of Pittsburgh, Pittsburgh, PA 15260, USA}

\author{Xiaowei Jia}
\affiliation{Department of Computer Science, The University of Pittsburgh, Pittsburgh, PA 15260, USA}

\author{Jens Eisert}
\email{jense@zedat.fu-berlin.de}
\affiliation{Dahlem Center for Complex Quantum Systems, Freie Universit{\"a}t Berlin, 14195 Berlin, Germany}

\author{Junyu Liu}
\email{junyuliu@pitt.edu}
\affiliation{Department of Computer Science, The University of Pittsburgh, Pittsburgh, PA 15260, USA}

\author{Jin-Peng Liu}
\email{liujinpeng@tsinghua.edu.cn}
\affiliation{Yau Mathematical Sciences Center and Department of Mathematics, Tsinghua University, Beijing 100084, China}
\affiliation{Yanqi Lake Beijing Institute of Mathematical Sciences and Applications, Beijing 100407, China}

\begin{abstract}
    A diffusion probabilistic model (DPM) is a generative model renowned for its ability to produce high-quality outputs in tasks such as image and audio generation. However, training DPMs on large, high-dimensional datasets such as high-resolution images or audio incurs significant computational, energy, and hardware costs. In this work, we introduce efficient quantum algorithms for implementing DPMs through various quantum ODE solvers. These algorithms highlight the potential of quantum Carleman linearization for diverse mathematical structures, leveraging state-of-the-art quantum linear system solvers (QLSS) or linear combination of Hamiltonian simulations (LCHS). Specifically, we focus on two approaches: DPM-solver-$k$ which employs exact $k$-th order derivatives to compute a polynomial approximation of $\epsilon_\theta(x_\lambda,\lambda)$; and UniPC which uses finite difference of $\epsilon_\theta(x_\lambda,\lambda)$ at different points $(x_{s_m}, \lambda_{s_m})$ to approximate higher-order derivatives. As such, this work represents one of the most direct and pragmatic applications of quantum algorithms to large-scale machine learning models, presumably taking substantial steps towards demonstrating the practical utility of quantum computing.
    \end{abstract}

\maketitle

\section{{Introduction}}\label{sec:introduction}

Diffusion models, in the 
context of machine learning and, more specifically, deep learning, represent a class of generative models that have garnered significant attention for their novel approach to modeling complex data distributions
\cite{DiffusionModels1,DiffusionModels2,DiffusionModels3,DiffusionModels3,ho2020denoising}. Unlike traditional models
of machine learning, which often rely on explicit likelihood estimation or adversarial training, diffusion models generate data through a gradual, iterative process inspired by the physical concept of diffusion. They 
work by initially introducing noise at data points and then learning how 
to reverse this process, effectively denoising the data step-by-step. This framework allows them to generate high-quality samples, often rivaling or surpassing the performance of other generative models, such as \emph{generative adversarial networks} (GANs) \cite{GANS} and \emph{variational autoencoders} (VAEs) \cite{VAE}. The appeal of diffusion models lies in their stability during training and their ability to model intricate, high-dimensional data distributions without requiring complex adversarial setups or delicate balancing of training dynamics. 
\emph{Diffusion probabilistic models} 
(DPM), as we call them, are families of generative
models renowned for their ability to produce 
high-quality out-
puts in complex tasks such as 
image and audio generation. 

Turning to quantum approaches, quantum computing is widely regarded as one of the most promising alternatives to von Neumann architectures in the post-Moore era \cite{preskill1998lecture}. By leveraging unique quantum features such as superposition and entanglement, quantum computing is anticipated to surpass classical counterparts in specific tasks. These include factoring \cite{shor1999polynomial}, database search \cite{grover1996fast}, simulating complex quantum systems \cite{lloyd1996universal}, sampling \cite{SupremacyReview}, and solving problems in linear algebra \cite{Wang_2024,Harrow_2009}.
Although we are currently still in what can be 
called the \emph{noisy intermediate-scale quantum} (NISQ) era, the concept of a \textit{megaquop machine} 
\cite{Megaquop} offers a vision of a future of \emph{fault-tolerant application-scale quantum} (FASQ) computing \cite{MindTheGaps} in which quantum computing achieves remarkable reliability and scalability. 
Such a megaquop machine, with an error rate per logical gate on the order 
of $10^{-6}$, would surpass the limitations of classical, NISQ, or analog quantum devices. 
It would have the capacity to execute quantum circuits involving around $100$ logical qubits and 
a circuit depth of 
approximately $10,000$—capabilities far beyond the reach of 
current technology.  

\begin{figure*}[t!]
    \centering
    \includegraphics[width=1\linewidth]{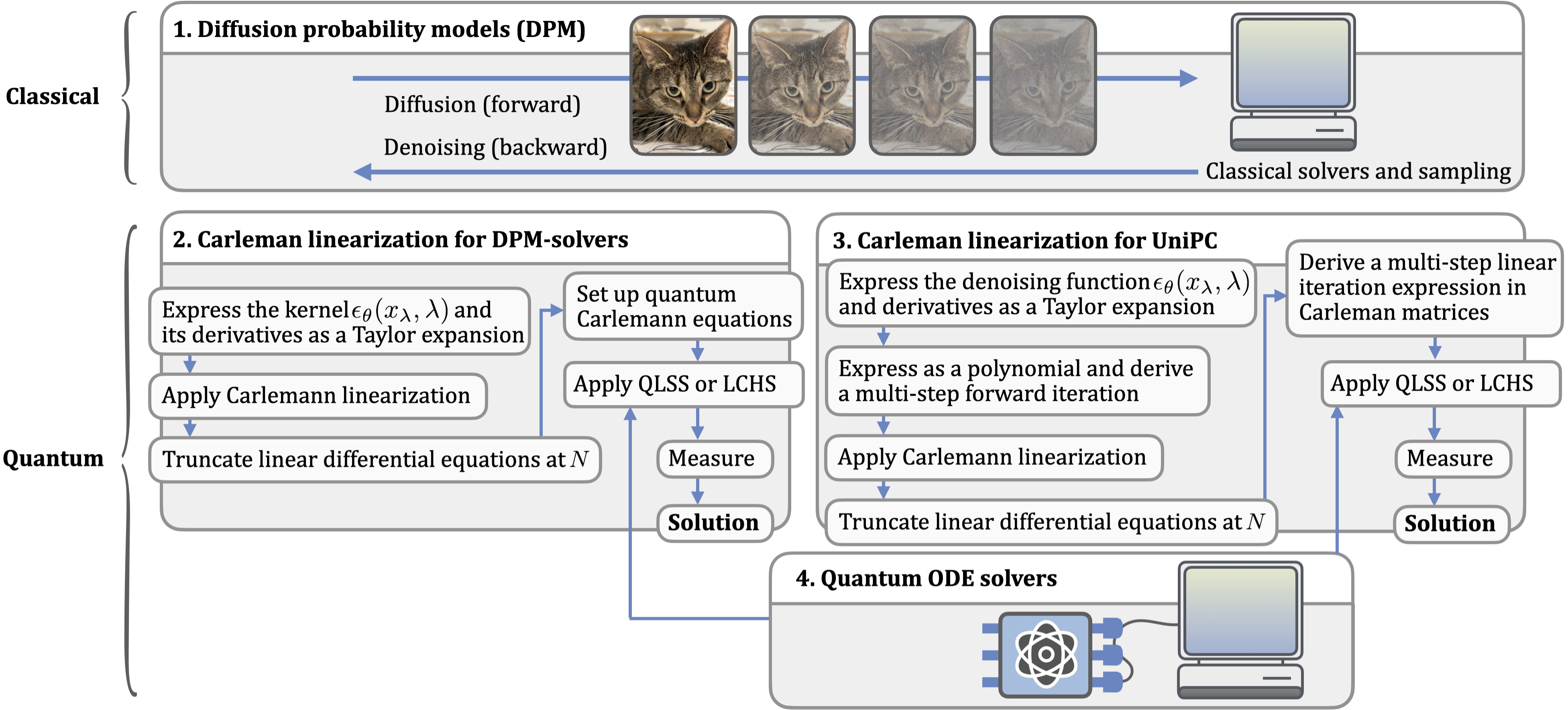}
    \caption{
    Sketch of classical (1.) \emph{diffusion probability 
    models} (DPM) and the 
    quantum algorithms proposed here for quantum analogs 
    (2.-4.). We pursue two approaches: 
   This is on 
   the one hand Carleman linearization for DPM solvers (2.), on the other 
    Carleman linearization for UniPC
    (3.). For both approaches, quantum ODE 
    solvers are
    instrumental in the step before measurement. 
    QLSS stands for \emph{quantum linear system solvers}, LCHS for \emph{linear 
    combination of Hamiltonian simulation}.
    } 
    \label{fig:BigFigure}
\end{figure*}

In the light of these observations, it is not surprising that  recent years have seen a surge of interest in exploring how quantum algorithms might enhance machine learning tasks, whether in terms of sample complexity, computational complexity, or generalization \cite{biamonte2017quantum,RevModPhys.91.045002,McClean_2016,Wang_2024}.
Indeed, this interest is well motivated by the fact that, for certain structured models, quantum algorithms are proven to vastly outperform their classical counterparts. In fact, rigorous separations have been 
demonstrated for specific PAC learning problems \cite{PACLearning,DensityModelling,TemmeML}. 
Even constant-depth quantum circuits have been shown to offer advantages over suitably constrained classical circuits \cite{ShortCircuitsLearning}. 
These insights provide strong motivation for the rapidly evolving field of quantum machine learning and give impetus for the thought that quantum computers may well offer computational advantages.

That said, quantum computers excel at tackling highly 
structured problems, while many common machine learning challenges involve highly unstructured data. In the light of this, it remains to 
an extent unclear what practical advantages quantum computers 
might offer for addressing machine learning tasks. Furthermore, it is still uncertain whether proving separations will be the 
most 
effective or even the best approach for advancing the field \cite{PRXQuantum.3.030101}.
For variational approaches \cite{McClean_2016}, 
while being exciting and promising,
no rigorous guarantees and separations of quantum 
over
classical approaches are known yet.
To date, it seems fair to say that  are lacking a framework for devising end-to-end applications of quantum computers for industrially relevant machine learning tasks \cite{Myths}.

In this work, 
we take out-of-the-box steps to establish a fresh avenue towards showing the utility of quantum computers in machine learning tasks.
Concretely, we present novel and efficient quantum algorithms designed for implementing \emph{differential path models} (DPMs) through a variety of quantum \emph{ordinary differential equation} (ODE) solvers. 
Our algorithms showcase the potential of quantum 
Carleman linearization in 
tackling a 
broad range of mathematical structures and problems, leveraging some of the latest advancements in \emph{quantum linear system solvers} 
(QLSS) or \emph{linear combination of Hamiltonian simulation} (LCHS). Specifically, we explore two distinctly
different but related approaches: the DPM-solver-$k$, which utilizes exact $k$-th order derivatives to compute a polynomial approximation of $\epsilon_\theta(x_\lambda, \lambda)$, and UniPC, which approximates higher-order derivatives by employing finite differences of $\epsilon_\theta(x_\lambda, \lambda)$ at multiple points $(x_{s_m}, \lambda_{s_m})$. Furthermore, this 
work represents one of the most direct and pragmatic applications of quantum algorithms to large-scale machine learning models, highlighting a key milestone in demonstrating the tangible benefits and practical utility of quantum computing in real-world tasks. We provide both rigorous statements and numerical results.
While this result may not yet be a full
end-to-end application, by
bridging the gap between theoretical quantum algorithms and their application to complex computational models, we take a significant step, so we think, toward realizing the power of quantum technologies for solving real-world problems in machine learning and beyond. 

\section{{Results}}\label{sec:results}

In this section, we lay out our specific results, both concerning rigorous statements that heuristic numerical statements that are meant to provide further evidence for the functioning of the approach. Substantially more detail on all steps will be presented in the supplemental material.
We will now explain what 
steps we take towards
establishing a fresh direction for demonstrating the utility of quantum computers in machine learning tasks. Specifically, we present novel and efficient quantum algorithms for implementing \emph{diffusion probabilistic models} (DPMs) using a variety of \emph{ordinary differential equation} (ODE) solvers. These algorithms leverage recent advancements in \emph{quantum ODE solvers} and highlight the potential of quantum Carleman linearization for addressing a broad range of mathematical structures and challenges.
Our exploration focuses on two distinct approaches.
\begin{itemize}
    \item DPM-solver-$k$: Utilizes exact $k$-th order derivatives to compute a polynomial approximation of $\epsilon_\theta(x_\lambda, \lambda)$.
    \item UniPC-$p$: Approximates higher-order derivatives by employing finite differences of $\epsilon_\theta(x_\lambda, \lambda)$ at different points $(x_{s_m}, \lambda_{s_m})$.
\end{itemize}
These algorithms underscore the 
potential of quantum Carleman linearization in addressing diverse mathematical structures. Moreover, as we advance toward the era of the era of FASQ \cite{MindTheGaps}, this work stands out as one of the most direct and practical applications of quantum algorithms to large-scale machine learning models. It marks, so we hope, a significant milestone in showcasing the tangible utility of quantum computing in real-world applications. 

\subsection{{Embedding classical neural networks}}

In the center of DPMs in the classical realm are ordinary differential equations capturing diffusion processes. On the highest level, they are defined by a \emph{forward process}, the reverse or \emph{backward} process, and the \emph{sampling procedure}.
In this work, the core aim is to establish quantum algorithms for quantum analogs of such models. Since quantum mechanics is intrinsically linear, much of the work will circle around embedding the problem in the appropriate fashion.
In this section, we present a brief heuristic overview of how to solve the diffusion ODE
\begin{equation}
    \frac{\mathrm{d} x_t}{\mathrm{d} t} = f(t) x_t + \frac{g^2(t)}{2\sigma_t} \epsilon_{\theta}(x_t,t), \qquad x_T \sim q_T(x_T),
\label{eq:diffusion-ODE}
\end{equation}
leveraging an instance of a quantum ODE solver based on both the DPM-solver \cite{lu2022dpm,lu2022dpm++} and UniPC \cite{zhao2024unipc}.

\subsubsection{{Carleman linearization for DPM-solvers}}
We here outline the core ideas of the linearization and embedding -- details can be found in the supplemental material, also shown in Fig.\ \ref{fig:BigFigure}.
The DPM-solver series are currently the most widely used solvers for diffusion ODEs. Given an initial value $x_s$, 
the solution $x_t$ of the diffusion ODE w.r.t.\ 
the noise prediction model Eq.~(\ref{eq:diffusion-ODE}) for $t \in (0,s)$ is given by
\begin{equation}
    \frac{x_t}{\alpha_t} = \frac{x_s}{\alpha_s} - \int_{\lambda_s}^{\lambda_t} e^{-\lambda} \epsilon_{\theta}(x_{\lambda},\lambda) \mathrm{d} \lambda.
\label{eq:diffusion-ODE-exact}
\end{equation}
In this expression $\lambda \coloneqq \log(\alpha_t/\sigma_t)$ is what is called the log-SNR variable. For a given a set of time steps  
$\{t_i\}_{i=0}^M$ with $t_0 = T$ and $t_M = 0$, with $h_i = \lambda_{t_i} - \lambda_{t_{i-1}}$, one makes use of the $(k-1)$-st-order Taylor approximation in order to develop
what is called the DPM-solver-$k$ in Ref.~\cite{lu2022dpm}. This actually takes the form
\begin{equation}\label{eq:DPM-solver-k}
    \begin{split}
        x_{t_i} &= \frac{\alpha_{t_i}}{\alpha_{t_{i-1}}}  x_{t_{i-1}}
        - \alpha_{t_i} \sum_{n=0}^{k-1} \epsilon^{(n)}_{\theta}(x_{\lambda_{t_{i-1}}},\lambda_{t_{i-1}})\\
        & \times \int_{\lambda_s}^{\lambda_t} e^{-\lambda} \frac{(\lambda - \lambda_{t_{i-1}})^n}{n!} \mathrm{d} \lambda  + O(h_i^{k+1}),
    \end{split}
\end{equation}
where $\epsilon^{(n)}_{\theta}(x_{\lambda_{t_{i-1}}},\lambda_{t_{i-1}})$ is the $n$-th derivative. 
The integral in the final line can 
be computed analytically in this case. In the same way, a polynomial expansion on the data prediction model $x_{\theta}(x_{\lambda},\lambda)$ gives rise to what is called the DPM-solver++~\cite{lu2022dpm++}. However, 
as mentioned before, 
the intrinsic linearity of quantum mechanics renders a direct implementation of non-linear ODE schemes inapplicable,
so that one has to 
resort to an instance of
Carleman or 
Koopman linearization and embedding of 
the diffusion ODE. In 
what follows, we discuss the logic and structure of the proposed quantum algorithm.
\begin{itemize}
    \item In a first step, we express $\epsilon_{\theta}(x_{\lambda},\lambda)$ and its higher-order derivatives as a polynomial expansion in $x_{\lambda}$. Building on this, we derive a forward iteration $x_{t_i} = \operatorname{poly}(x_{t_{i-1}})$. 
    
    \item For this still being a non-linear expression, we employ 
     \textit{Carleman linearization} to transform it into a linear system, so that a quantum algorithm can be devised.

    \item For this, we define $y_{j}(\lambda) \approx x^j_{\lambda}$ for $j = 1, \dots, N$. As the Carleman linearization transforms a non-linear differential equation into an infinite-dimensional system of linear differential equations, we truncate it at step $N$, with the initial condition $y_j(\lambda_{t_{i-1}}) = x^j_{\lambda_{t_{i-1}}}$.

    \item Setting $Y = (y_1, \ldots, y_N)$, we then derive a linear iteration of the form 
    \begin{equation} \label{LinearizedSystemDPM}
        \begin{split}
            \delta \hat{Y} & = \hat{Y}(t_{i}) - \hat{Y}(t_{i-1}) = A \hat{Y}(t_{i-1}) + b~,\\
            & \text{ for }\quad \hat{y}_1(t_M) = \hat{y}_1(0) = x_{\mathrm{in}}(0)~,
        \end{split}
    \end{equation}
    based on Eq.~(\ref{eq:DPM-solver-k}), introducing a matrix $A$, which is the \textit{quantum Carleman matrix} 
    (QCM) associated with the DPM-solver.

\end{itemize}
Eq.~(\ref{LinearizedSystemDPM}) embodies
 $M+1$ iterations in total ranging from $t_0 = T$ to $t_M = 0$. Here, $X_{\text{in}} = X(0)$ is specified by the initial vector. This
 expression, distinctly departing from that of 
 Ref.~\cite{Liu:2023coc}, has been designed to reconstruct the original data $x_0$ from $x_t$ through a \textit{backward} denoising diffusion process, contrasting with the forward progression typically seen in a (stochastic) gradient descent algorithm during the training process. 

\subsubsection{{Carleman linearization for UniPC}}

Complementing the above methodology, in this subsection, we design
a reading of a quantum Carleman algorithm suitable for the known 
alternative technique to solve the diffusion ODE referred to as UniPC. This includes both the predictor (UniP-p) and the corrector (UniC-p), defined as
\begin{equation}
    \begin{split}
        x_{s_p} = \frac{\alpha_{s_p}}{\alpha_{s_0}}x_{s_0} & - \sigma_{s_p} (e^{h_i} - 1) \epsilon_{\theta}(x_{s_0},\lambda_{s_0})\\
        & - \sigma_{s_p}B(h_i) \sum_{m=1}^{p-1} \frac{a_m}{r_m} D_m,
    \end{split}
\label{eq:UniP--p}
\end{equation}
and as
\begin{equation}
    \begin{split}
        x^c_{s_p} = \frac{\alpha_{s_p}}{\alpha_{s_0}}x^c_{s_0}& - \sigma_{s_p} (e^{h_i} - 1) \epsilon_{\theta}(x_{s_0},\lambda_{s_0})\\
        & - \sigma_{s_p}B(h_i) \sum_{m=1}^p \frac{a_m}{r_m} D_m.
    \end{split}
\label{eq:UniC--p}
\end{equation}
Here, we have encountered the interpolation time steps $s_m = h_ir_m + \lambda_{t_{i-1}}$, where $m = 1, 2, \dots, p$, and the finite difference $D_m = \epsilon_{\theta}(x_{s_m},\lambda_{s_m}) - \epsilon_{\theta}(x_{s_0},\lambda_{s_0})$. 
Note that
in this expression $s_0 = \lambda_{t_{i-1}}$, and $s_p = \lambda_{t_i}$.
Again, we outline the logic and structure of the proposed quantum algorithm.
\begin{itemize}
    \item In a first step, as before, we rewrite the denoising function $\epsilon_{\theta}(x_{\lambda},\lambda)$ in terms of a polynomial expansion in $x_\lambda$
    up to order $J$, 
    with the coefficients $a_j$ contains $\lambda$. 

    \item In a next step, we can express $\epsilon_{\theta}(x_{\lambda},\lambda)$ at different points as a polynomial of $x_{\lambda}$, and derive a multi-step 
    forward iteration for UniP given by $x_{s_p} = \operatorname{poly}(x_{s_0}) + \operatorname{poly}(x_{s_1}) + \cdots + \operatorname{poly}(x_{s_{p-1}})$. For UniC, this is $x^c_{s_p} = x^c_{s_0} + \operatorname{poly}(x_{s_0}) + \operatorname{poly}(x_{s_1}) + \cdots + \operatorname{poly}(x_{s_p})$.

    \item Similarly as before, we use \textit{Carleman linearization} in 
    this 
    step and truncate the system at $N$. We denote $y_j(\lambda) \approx x^j_{\lambda}(\lambda)$, $z_j(\lambda) \approx (x^c)^j_{\lambda}(\lambda)$ for $j = 1, \dots, N$, with the initial condition $y_1(s_0) = x_{\lambda}(s_0)$, $z_1(s_0) = (x^c)_{\lambda}(s_0)$. 

    \item In terms of $Y = (y_1, \ldots, y_N)$, and $Z = (z_1, \ldots, z_N)$, we can build on Eq.~(\ref{eq:UniP--p}) and Eq.~(\ref{eq:UniC--p}) 
    to derive a multi-step linear iteration (predictor) expression 
    given by \begin{equation}\label{LinearDifferenceEqnUniP}
        \begin{split}
            \delta Y(t_{i}) =~& A^{(0)} Y^{(0)}(t_{i}) + b^{(0)}\\
            & + A^{(1)} Y^{(1)}(t_{i}) + b^{(1)}\\
            & + \ldots + A^{(p-1)} Y^{(p-1)}(t_{i}) + b^{(p-1)}~,
         \end{split}
    \end{equation}
    with Carleman matrices $A^{(0)}, \dots, A^{(p-1)}$ for all 
    interpolation 
    steps. 
    The corrector, reflecting the other linear iteration
    takes the form
    \begin{equation} \label{eq:UniPCDifferenceEquation}
        \begin{split}
            \delta Z(t_i) = B Z(t_{i}) & + A^{(0)_c} Y^{(0)}(t_{i}) + b^{(0)_c}\\
            & + A^{(1)_c} Y^{(1)}(t_{i}) + b^{(1)_c}\\
            & + \ldots + A^{(p)_c} Y^{(p)}(t_{i}) + b^{(p)_c}~,
        \end{split}
    \end{equation} 
    with Carleman matrices $B, A^{(0)_c}, \dots, A^{(p)_c}$.

\end{itemize}
It is important to stress that 
the Carleman linearization approach 
introduced here represents a novel Carleman predictor-corrector 
scheme. This approach differs from both the 
DPM-solver framework described in the previous section and the approach taken in Ref.~\cite{Liu:2023coc}. A defining feature of the quantum Carleman algorithm for UniPC is the generation of a unique \emph{quantum Carleman matrix} (QCM) for each interpolation time step $s_m$ that we think is also interesting as an innovative step in its own right.

A practical limitation of the quantum Carleman linearization lies in the truncation of the polynomial hierarchy, which introduces a systematic truncation error that is difficult to control analytically. The contribution of the neglected higher-order terms generally depends on the nonlinearity strength and on how strongly probability amplitudes populate higher moments during evolution \cite{liu2021efficient}. In complex diffusion models, this error may grow over time or under strong coupling, and we do not yet have a rigorous bound for it. We emphasize this as an open technical limitation rather than a hidden assumption.

That said, for many physically relevant regimes—particularly when $N$ is large and the diffusion dynamics remains near a stable manifold—the effect of higher-order monomials is expected to be suppressed. Empirically, for smooth drifts and moderate noise levels, low-order truncations reproduce accurate dynamics because higher-order correlations decay rapidly or remain small. Thus, although the truncation error is not strictly controllable, it can remain practically negligible within the accuracy relevant for the applications studied here. A full characterization of this error scaling and its dependence on truncation order $N$ and diffusion strength, and truncation order is left for future work.

\begin{figure*}[t!]
    \centering
    \includegraphics[width=0.95\linewidth]{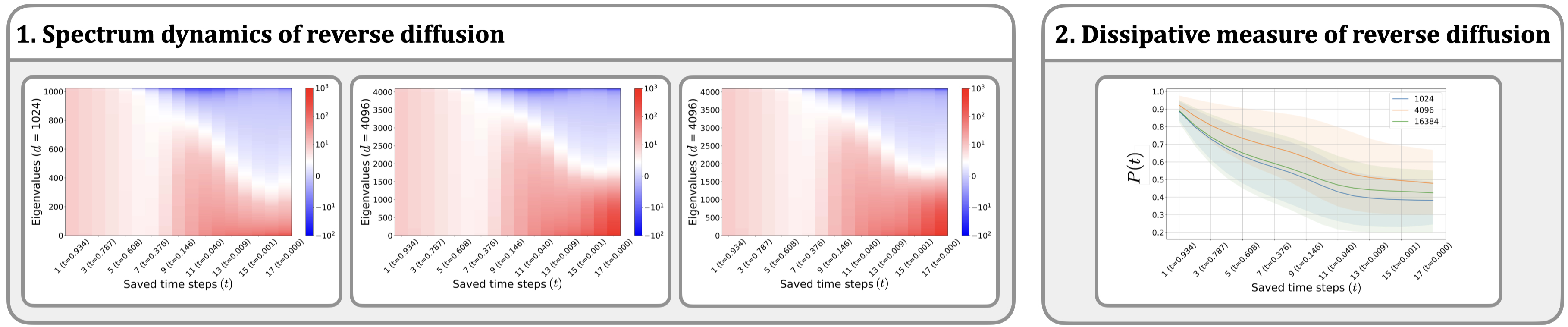}
    \caption{Results of numerical experiments.
    1. \textit{Spectrum dynamics of the reverse diffusion process on ImageNet-100.}
Using the DPM-Solver \citep{lu2022dpm}, we show the eigenvalue trajectories of the Jacobian 
$\mathbf{J}(t)$ for dimensions $d=1024,4096$, and $16384$, each representing image generation at a different resolution, ranging from coarse to fine-grained. Across all three cases, the eigenvalues start positive and progressively develop a larger spectral gap with extremely positive values, highlighting the system’s consistent dissipative behavior.
2. \textit{Dissipative measure of the reverse diffusion process on ImageNet-100.}
The shaded area represents the [25th, 75th] percentile of the statistics. Across three models with $d=$ 1024, 4096, and 16384, $P(t)$ consistently decreases with $t$.
    } 
    \label{fig:BigFigure2}
\end{figure*}

\subsection{{End-to-end quantum ODE solver}}



With the foundational elements established, there are two distinct approaches to solving the resulting linear ODEs. The first approach leverages the advanced capabilities of the \emph{quantum linear system solver} (QLSS), while the second utilizes the \emph{linear combination of Hamiltonian simulation} (LCHS) algorithm. We will begin by exploring the application of QLSS and subsequently discuss the LCHS method.

To make use of QLSS, we need to massage the iteration ODEs into a single vector equation, as has been done in Ref.~\cite{Liu:2023coc}. 
We look at the DPM-solver first, the entire
iteration scheme established in Eq.~(\ref{LinearizedSystemDPM}) can be captured in a vector equation 
\begin{equation}\label{LinearSystemDPMSolver}
    M \boldsymbol{Y} = \boldsymbol{b}~
\end{equation}
that collects $Y(t_i)$ at all time steps in order to form a single big vector $\boldsymbol{Y} = (Y(t_0), Y(t_1), \dots, Y(t_M))$. Similarly, the UniPC system can also be captured in a single vector equation as
\begin{equation}\label{UniPCLinearSys}
    \boldsymbol{\mathsf{M}^c} \boldsymbol{\mathsf{Z}} = \boldsymbol{\mathsf{b}^c}~,
\end{equation}
where $\boldsymbol{\mathsf{Z}}$ collects all of the $Z(t_i)$. The explicit forms of Eq.~(\ref{LinearSystemDPMSolver}) and Eq.~(\ref{UniPCLinearSys}) are provided in the supplementary 
material.

We are 
now in the position to invoke 
and discuss an instance of a \emph{quantum linear system solver} (QLSS) to efficiently solve the 
resulting differential equations. QLSS have technically advanced significantly since the introduction of the original \emph{Harrow-Hassidim-Lloyd} (HHL) algorithm proposed in 2008, which has achieved a complexity of $\mathcal{O}(\operatorname{poly}(\kappa, 1/\epsilon))$ for solving sparse linear systems, where $\kappa>0$ is the condition number and $\epsilon>0$ is the desired accuracy \cite{Harrow_2009}. Subsequent developments, including Ambainis' \emph{variable-time amplitude amplification} (VTAA) and linear combination of quantum walks,
have improved the scaling to $\mathcal{O}(\kappa \log(1/\epsilon))$ \cite{ambainis2010variable, Childs_2017}. These algorithms utilize matrix ($\mathcal{O}_M$) and state preparation ($\mathcal{O}_{\boldsymbol{b}}$) oracles, with $\mathcal{O}_{\boldsymbol{b}}$ often incurring lower costs 
in practice \cite{low2024QLAOptimal}. Further recently introduced techniques, such as kernel reflection and adiabatic evolution, have further optimized 
the query complexities, achieving near-optimal bounds for both $\mathcal{O}_M$ and $\mathcal{O}_{\boldsymbol{b}}$ \cite{costa2021optimal, dalzell2024shortcut}. These advancements are particularly relevant for sparse systems like those arising in the Carleman method, where the matrix sparsity $s_M$ and the success probability $p_{\text{succ}}$ significantly influence the overall complexity \cite{Gily_n_2019}. 

Going further, 
and building on the
theorems and quantum algorithms discussed 
and introduced in Refs.~\cite{Liu:2023coc,costa2021optimal,dalzell2024shortcut,low2024QLAOptimal}, we derive the formal version of Theorem \ref{DPMsolverComplexityTheorem} and Theorem \ref{UniPCComplexityTheorem} presented in the next section, which establish the gate complexity bounds for the quantum Carleman algorithm (for the formal version of these theorems, we refer to the supplementary material). These results integrate advancements in quantum linear system solvers, particularly leveraging the state-of-the-art methods for efficient matrix block encoding, initial state preparation, and sparsity-aware optimization, ensuring a comprehensive understanding of the computational costs associated with solving such systems.

Besides the QLSS approach, the \emph{linear combination of Hamiltonian simulation} (LCHS) strategy \cite{An_2023,an2023quantum} provides an alternative method for solving the linear ODEs in Eqs.~(\ref{LinearSystemDPMSolver}) and (\ref{UniPCLinearSys}) by representing the solution to the homogeneous equation as a linear combination of Hamiltonian simulations. When a source term (e.g., $\boldsymbol{\mathsf{b}^c}$ in Eq.~(\ref{UniPCLinearSys}) and the right-hand side of Eq.~(\ref{LinearSystemDPMSolver})) is present, Duhamel's principle 
can be applied \cite{An_2023}. LCHS achieves gate efficiency by combining \emph{linear combinations of unitaries} (LCU) \cite{Childs2012Hamiltonian} with existing 
Hamiltonian simulation techniques. The method involves truncating the infinite integral to a finite interval $[-K,K]$ and discretizing it using the  $M + 1$ grid points. Compared to the standard QLSS approach, LCHS directly implements the time evolution operator, significantly reducing the number of state preparation oracles required.

At last, we address briefly the challenges of uploading and downloading data for our quantum algorithms, as this aspect is sometimes underappreciated. By leveraging sparsity in both data representation and training, our framework \emph{avoids} the need for \emph{quantum random access memory}  (QRAM) \cite{Liu:2023coc}, thus enhancing scalability and practicality while minimizing classical-to-quantum communication overhead \cite{hann2019hardware, Matteo_2020, Hann_2021, Wang:2023oon}. 
The downloading problem is actually more challenging than uploading, as it involves performing state tomography \cite{BenchmarkingReview} on the resulting quantum states from our quantum algorithms \cite{aaronson2018shadowtomography,Huang_2020,Wang_2024}. We focus on tomographic recovery in sparse training on sparse vectors and QCM matrices from the DPM architecture, which is similar to the structure presented in Ref.~\cite{Liu:2023coc}. The recovery process identifies $r$-sparse computational basis vectors and reconstructs the state using a modified \emph{classical shadow estimation} scheme with $n$-qubit Clifford circuits \cite{Huang_2020,ShallowShadows,BenchmarkingReview}, requiring $\mathcal{O}(r \log r)$ measurements only. 

\subsection{{Theorems}}

The above scheme give rise to practical quantum algorithms for a fault tolerant quantum computer.
In this section, we will lay out the informally formulated main theorems that are established in this work. Details can be found in the supplementary material. 

\begin{theorem}[Informal complexity of the DPM-solver] \label{DPMsolverComplexityTheorem}
    For a DPM-solver with a denoising function expanded up to the $J$-th order Taylor approximation and truncated at $N$, the system of linear equations $M \boldsymbol{Y} = \boldsymbol{b}$ can be solved using a quantum algorithm. If $M$ is sparse and block-encoded efficiently, the quantum state proportional to the solution vector can be prepared with query complexity scaling as
    \begin{equation}
        \mathcal{O}(J \kappa \cdot \mathrm{polylog}(N/\epsilon))~,
    \end{equation}
    where $\kappa>0$ is an upper bound on the condition number of $M$. The algorithm achieves a success probability $p_{\text{succ}}$ greater than $1/2$ and accuracy $\epsilon>0$.

    Assuming the output weight vectors are \( r \)-sparse with \( m \coloneqq \log_2(N) \), the tomographic cost for transferring quantum states to classical devices is \( \mathcal{O}(m^2 r^3 /\epsilon^2) \), ignoring logarithmic factors. The algorithm excludes the state preparation cost, which remains efficient for sparse training.
\end{theorem}

The next statement summarizes the sample complexity of the second approach taken.

\begin{theorem}[Informal complexity of the UniPC-$p$ framework]\label{UniPCComplexityTheorem}
    For the UniPC-$p$ framework used in fast sampling of DPMs, a system of linear equations $\boldsymbol{\mathsf{M}^c} \boldsymbol{\mathsf{Z}} = \boldsymbol{\mathsf{b}^c}$ can be formulated and solved using a quantum algorithm. If $\boldsymbol{\mathsf{M}^c}$ is sparse and efficiently block-encoded, the quantum state proportional to the solution can be prepared with query complexity 
    \begin{equation}
        \mathcal{O}(p J \kappa \cdot \mathrm{polylog}(N/\epsilon))~,
    \end{equation}
    where $\kappa>0$ bounds the condition number of $\boldsymbol{\mathsf{M}^c}$. The algorithm achieves a success probability $p_{\text{succ}}$ greater than $1/2$ and accuracy $\epsilon>0$.

    Assuming the output weight vectors are \( r \)-sparse with \( m \coloneqq \log_2(N) \), the tomographic cost for transferring quantum states to classical devices is \( \mathcal{O}(m^2 r^3 /\epsilon^2) \), ignoring logarithmic factors. The algorithm excludes the state preparation cost, which remains efficient for sparse training.
\end{theorem}
These statements ensure that the quantum algorithms are efficiently implementable on a fault tolerant quantum computer.


\subsection{{Numerical analysis and experiments}}\label{sec:experiments}

Above, we present the performance guarantees of the quantum algorithms
introduced. It is a core contribution of this work, however, to also make the 
point that the approach taken is reasonably applicable to real-world data.
This is a valuable contribution, as the proven separations for machine learning
problems are for highly structured and artificial data \cite{PACLearning,DensityModelling,TemmeML}. In this section, we perform
extensive numerical experiments to show that for real-world
datasets, the conditions of the above theorems are commonly satisfied for natural data
and that the processes considered are sufficiently diffusive in a precise
sense. 

%
In order to do so, 
we employ U-ViT \citep{bao2022all} operating in the latent space of a pre-trained AutoEncoder from Stable diffusion \citep{rombach2021highresolution}. Working on the ImageNet-100 dataset \citep{deng2009imagenet}, the AutoEncoder compresses input images into a lower-dimensional latent space, reducing computational overhead while preserving perceptual quality. The forward diffusion process follows the variance-preserving schedule \citep{DiffusionModels3}. We train the U-ViT-M for 200 epochs using the AdamW optimizer.  
During sampling, we employ the DPM-Solver-fast, a mix of single-order solvers \citep{lu2022dpm}. 
Specifically, it uses 17 pre-defined saved time steps to reverse the diffusion trajectory, with each step involving intermediate substeps to approximate the continuous ODE solution, resulting in 50 \emph{number of function evaluations} (NFEs). 

To evaluate the dissipative property of the reverse diffusion process, we 
explore whether trajectories contract in phase space over time. This involves analyzing the system's spectral properties by computing the Jacobian of its drift term as
\begin{equation}
    \boldsymbol{J}(t) = f(t) + \frac{g^2(t)}{2\sigma_t}\frac{\partial  \epsilon_{\theta}(x_t,t)}{\partial x_t},
\end{equation}
where we further compute the eigenvalues $\lambda_i, i=1,\dots, d$, of its symmetric form $\boldsymbol{J}(t) + \boldsymbol{J}(t)^\dagger$.
We visualize the spectral dynamics of the sampling process for well-trained diffusion models with different dimensionality.
Specifically, we project images with resolutions $128\times128$, $256\times256$, and $512\times512$ into the latent space of $16\times16\times4$, $32\times32\times4$, and $64\times64\times4$.
In the light of the above rigorous statements, 
a quantum advantage is plausible if dissipative
modes exist which are 
characterized by the positivity of the Hessian eigenvalues in Ref.\ \cite{Liu:2023coc}. For the reverse-time diffusion, this amounts to testing the positivity of 
the eigenvalues of the Jacobian.
As shown in Fig.~\ref{fig:BigFigure2} 1.,
the sampling processes at all three resolutions exhibit a consistent pattern: They begin with all positive eigenvalues and gradually transition to a state where the spectral gap increases significantly, with some eigenvalues becoming extremely positive. This 
growing spectral gap suggests stronger dissipativity in the system, where certain directions in the phase space contract more rapidly than others.
To further verify the trend of the eigenvalues, we define $P(t)$ as a function of normalized eigenvalues as
\begin{equation}
P(t) \coloneqq \frac{1}{d}\sum_{i=1}^d \Pi_{t'=T}^t (1-a_i(t')),
\end{equation}
where $a_i(t) \coloneqq {\lambda_i(t)}/{\max_{i,t} |\lambda_i(t)|}$.
By construction, $t\mapsto P(t)$ grows unboundedly when negative eigenvalues dominate and decays towards zero when positive eigenvalues prevail.
As shown in Fig.~\ref{fig:BigFigure2} 2., across different dimensionalities of the state variables, the inference process follows a consistent pattern: $P(t)$ steadily decreases as $t$ decreases, implying the emergence of large eigenvalues along the generative process, reinforcing the existence of dissipative modes, as it is required in the assumptions of the theorems.

\section{{Discussion}}\label{sec:concoutlook}

In this work, we have taken steps towards identifying practical applications of quantum computers in realistic machine learning tasks, by formulating meaningful analogs of diffusion probabilistic models that constitute an increasingly important family of classical generative models, producing high quality outputs. 
We build on and further develop tools of quantum solvers for ordinary differential equations to make progress in presenting 
quantum algorithms for quantum analogs of DPMs.
Our work showcases the potential of quantum Carleman linearization for a wide range of mathematical structures, utilizing cutting-edge quantum linear system solvers and linear combination of Hamiltonian simulations. We explain the respective features of the two approaches taken. For our rigorous results, we present performance guarantees, while we offer also results of numerical experiments to provide further evidence for the functioning of the approach.

At a broader level, we aim for our work to help address a key bottleneck in exploring the potential of quantum computing in machine learning. While quantum computers have demonstrated advantages for certain highly structured problems, there is a prevailing view that they may offer limited benefits for tasks in machine learning, which typically involve less structured data and more noise
\cite{PRXQuantum.3.030101}. This work, suggests, however, that there may be significant potential for quantum computers in performing inference tasks or generative tasks of this nature.
What is more, the present ansatz suggests that the
quantum advantage is not only present for highly engineered instances, but in fact \emph{robust} for an entire class of meaningful instances \cite{MindTheGaps}. Considering that inference costs might be more than training costs in many machine learning applications, our proposed approach could pave the way for a practical example of quantum advantage in machine learning industry, while also providing a pathway to making quantum computers more useful for solving industrially relevant problems.\\

\section*{{Acknowledgements}}\label{sec:ack}

J.~E.~has been supported by the BMFTR (Hybrid++, DAQC, QSolid), the QuantERA (HQCC),
the BMWK (EniQmA), the DFG (CRC 183), the Quantum Flagship (PasQuans2, Millenion), the Munich Quantum Valley, 
Berlin Quantum, and the European Research Council (DebuQC). 
J.-P.~L.~acknowledges support from Tsinghua University and Beijing Institute of Mathematical Sciences and Applications.
\newpage
\pagebreak
\clearpage
\foreach \x in {1,...,\the\pdflastximagepages}
{
	\clearpage
	

\section*{Supplementary material}

# Supplemental material: Towards efficient quantum algorithms for diffusion probability models

## I. DIFFUSION MODELS VIA ODE SOLVERS

*Diffusion probabilistic models* (DPMs) [1–4] are emerging as powerful generative models with remarkable performance across various tasks, including image generation, video generation, text-to-image generation, speech synthesis, and lossless compression [5]. DPMs are defined either by discrete-time random processes [4] or continuous-time *stochastic differential equations* (SDEs) [3], which learn to progressively denoise data points.

### A. Diffusion (probability flow) ordinary differential equations

Suppose we have a $D$-dimensional data $x_0 \in \mathbb{R}^D$ vector, DPMs define a *forward diffusion process* $\{q_t\}_{t=0}^T$ to gradually add Gaussian noise to degenerate the unknown data distribution $q_0(x_0)$ to an approximately Gaussian distribution $q_T(x_T) \approx \mathcal{N}(x_T \mid 0, \tilde{\sigma}^2)$ for some $\tilde{\sigma} > 0$, where the transition kernel satisfies

$$q_{0t}(x_t|x_0) = \mathcal{N}(x_t|\alpha_t x_0, \sigma_t^2 \boldsymbol{I}). \tag{I.1}$$

Here $\alpha_t, \sigma_t \in \mathbb{R}^+$, are differentiable functions of $t$ with bounded derivatives, and we denote them as $\alpha_t, \sigma_t$ for simplicity. The choice for $\alpha_t$ and $\sigma_t$ is referred to as the *noise schedule* of a DPM. The *signal-to-noise-ratio* (SNR) $\alpha_t^2/\sigma_t^2$ is strictly decreasing with respect to $t$ [5–7]. The transition can be rewritten as $x_t = \alpha_t x_0 + \sigma_t^2 \epsilon_t$, where $\epsilon_t \sim \mathcal{N}(0, \boldsymbol{I})$. Equivalently, the *forward noising diffusion*

$$dx_t = f(t)x_t dt + g(t)dw_t, \qquad x_0 \sim q_0(x_0), \tag{I.2}$$

where $w_t \in \mathbb{R}^D$ is the standard Wiener process [7], and

$$f(t) = \frac{d \log \alpha_t}{dt}, \quad g^2(t) = \frac{d\sigma_t^2}{dt} - 2\frac{d \log \alpha_t}{dt}\sigma_t^2. \tag{I.3}$$

DPMs learn to recover the data $x_0$ from $x_t$ with a backward denoising diffusion process. Song et al. [3] have shown that the forward process Eq. (I.2) can be associated with a reverse-time process from $T$ to 0, named as the *backward denoising diffusion*

$$dx_t = [f(t)x_t - g^2(t)\nabla_x \log(q_t(x_t))]dt + g(t)d\bar{w}_t, \qquad x_T \sim q_T(x_T), \tag{I.4}$$

where $\bar{w}_t$ is a standard Wiener process in the reverse time. The basic idea of DPMs is to estimate the *score function* $-\sigma_t \nabla_x \log(q_t(x_t))$ by a *time-dependent neural network* $\epsilon_\theta(x_t, t)$ parameterized by $\theta$, known as the *score network*. As $\epsilon_\theta(x_t, t)$ can also be regarded as predicting the Gaussian noise added to $x_t$, it is also known as the *noise prediction model* [3, 4].

Alternatively, the *data prediction model* $x_\theta(x_t, t)$ estimate the data $x_0$ based on the noisy $x_t$, and its relation with $\epsilon_\theta(x_t, t)$ is given by $x_\theta(x_t, t) = (x_t - \sigma_t \epsilon_\theta(x_t, t))/\alpha_t$ [7], or $\epsilon_\theta(x_t, t) = (x_t - \alpha_t x_\theta(x_t, t))/\sigma_t$. It is derived from the exact relation $-\sigma_t \nabla_x \log(q_t(x_t)) = (x_t - \alpha_t x_0)/\sigma_t$. DPMs substitute the score function $-\sigma_t \nabla_x \log(q_t(x_t))$ with the noise prediction model $\epsilon_\theta(x_t, t)$ in Eq. (I.4), giving a parameterized reversible process, the (*diffusion SDE*)

$$dx_t = \left[f(t)x_t + \frac{g^2(t)}{\sigma_t}\epsilon_\theta(x_t, t)\right]dt + g(t)d\bar{w}_t, \qquad x_T \sim \mathcal{N}(\boldsymbol{0}, \tilde{\sigma}^2 \boldsymbol{I}). \tag{I.5}$$

Numerical SDE solvers suffer from the the randomness of the Wiener process [5]. For faster sampling, Song et al. [3] have proven that there is an alternative deterministic process

$$\frac{dx_t}{dt} = f(t)x_t - \frac{1}{2}g^2(t)\nabla_x \log(q_t(x_t)), \qquad x_T \sim q_T(x_T) \tag{I.6}$$

whose marginal distribution of $x_t$ is also $q_t(x_t)$. Similarly, by replacing $-\sigma_t \nabla_x \log(q_t(x_t))$ with $\epsilon_\theta(x_t, t)$ in Eq. (I.6), we obtain the *probability flow ODE (diffusion ODE)*

$$\frac{dx_t}{dt} = f(t)x_t + \frac{g^2(t)}{2\sigma_t}\epsilon_\theta(x_t, t), \qquad x_T \sim q_T(x_T). \tag{I.7}$$

The equivalent diffusion ODE w.r.t. the data prediction model $x_\theta(x_t, t)$ is given by

$$\frac{dx_t}{dt} = \left[f(t) + \frac{g^2(t)}{2\sigma_t^2}\right]x_t - \frac{\alpha_t g^2(t)}{2\sigma_t^2}x_\theta(x_t, t), \qquad x_T \sim q_T(x_T). \tag{I.8}$$

## B. Exponential integrators and DPM-solvers

There are a variety of ODE solvers for implementing the *probability flow ODE* (diffusion ODE) [5, 6, 8–11]. DPM-solver series are the currently most popular diffusion ODE solvers. Given an initial value $x_s$, the solution $x_t$ of the diffusion ODE w.r.t. the noise prediction model Eq. (I.7) ($t \in (0, s)$) is

$$\frac{x_t}{\alpha_t} = \frac{x_s}{\alpha_s} - \int_{\lambda_s}^{\lambda_t} e^{-\lambda} \epsilon_\theta(x_\lambda, \lambda) \mathrm{d}\lambda, \tag{I.9}$$

and the solution $x_t$ of the diffusion ODE w.r.t. the data prediction model Eq. (I.8) ($t \in (0, s)$) is

$$\frac{x_t}{\sigma_t} = \frac{x_s}{\sigma_s} + \int_{\lambda_s}^{\lambda_t} e^{-\lambda} x_\theta(x_\lambda, \lambda) \mathrm{d}\lambda. \tag{I.10}$$

Here $\lambda := \log(\alpha_t/\sigma_t)$ is the log-SNR variable. Given a set of decreasing time steps $\{t_i\}_{i=0}^M$ with $t_0 = T$ and $t_M = 0$, with $h_i = \lambda_{t_i} - \lambda_{t_{i-1}}$, the first-order Taylor expansion on $\epsilon_\theta(x_\lambda, \lambda)$ gives the DPM-solver-1 in Ref. [5]

$$x_{t_i} = \frac{\alpha_{t_i}}{\alpha_{t_{i-1}}} x_{t_{i-1}} - \sigma_{t_i}(e^{h_i} - 1) \epsilon_\theta(x_{t_{i-1}}, \lambda_{t_{i-1}}). \tag{I.11}$$

The $k$th-order Taylor approximation is then used to develop the DPM-solver-$k$ in Ref. [5]. The $(k-1)$-th order Taylor expansion of $\epsilon_\theta(x_\lambda, \lambda)$ w.r.t. $\lambda$ at $\lambda_{t_{i-1}}$ gives

$$\epsilon_\theta(x_\lambda, \lambda) = \sum_{n=0}^{k-1} \frac{(\lambda - \lambda_{t_{i-1}})^n}{n!} \epsilon_\theta^{(n)}(x_{\lambda_{t_{i-1}}}, \lambda_{t_{i-1}}) + O(h_i^k), \tag{I.12}$$

where $\epsilon_\theta^{(n)}(x_{\lambda_{t_{i-1}}}, \lambda_{t_{i-1}})$ is the $n$-th derivative. The DPM-solver-$k$ has the form [5]

$$x_{t_i} = \frac{\alpha_{t_i}}{\alpha_{t_{i-1}}} x_{t_{i-1}} - \alpha_{t_i} \sum_{n=0}^{k-1} \epsilon_\theta^{(n)}(x_{\lambda_{t_{i-1}}}, \lambda_{t_{i-1}}) \int_{\lambda_s}^{\lambda_t} e^{-\lambda} \frac{(\lambda - \lambda_{t_{i-1}})^n}{n!} \mathrm{d}\lambda + O(h_i^{k+1}). \tag{I.13}$$

The integral $\int_{\lambda_s}^{\lambda_t} e^{-\lambda}(\lambda - \lambda_{t_{i-1}})^n/n!$ can here be analytically computed. Essentially, we use a high-order forward ODE solver to compute $x_{t_i}$ by using high-order derivatives $\epsilon_\theta^{(n)}(x_{\lambda_{t_{i-1}}}, \lambda_{t_{i-1}})$. Similarly, a Taylor expansion on the data prediction model $x_\theta(x_\lambda, \lambda)$ gives rise to the DPM-solver++ [6]. However, the intrinsic linerity of quantum mechanics prevents a direct implementation of non-linear ODE schemes, requiring Carleman/Koopman linearization of the diffusion ODE.

## C. Unified predictor-corrector

As another approach for practically solving the diffusion ODE, we consider a state-of-the-art fast sampling method of DPMs: This is the so-called *unified predictor-corrector framework* (UniPC) [10]. This works as follows: Given a set of decreasing time $\{t_i\}_{i=0}^M$ with $t_0 = T$ and $t_M = 0$, with $h_i = \lambda_{t_i} - \lambda_{t_{i-1}}$. We also use a non-zero increasing schedule $r_1 < \cdots < r_p = 1$ to interpolate between $\lambda_{t_{i-1}}$ and $\lambda_{t_i}$ to obtain the auxiliary time steps: $s_m = h_i r_m + \lambda_{t_{i-1}}$ with $m = 1, 2, \ldots, p$. The UniPC-p combines the predictor (UniP-p) to estimate $x_{t_i}$ with the previous $p$ points according to

$$x_{t_i} = \frac{\alpha_{t_i}}{\alpha_{t_{i-1}}} x_{t_{i-1}} - \sigma_{t_i}(e^{h_i} - 1) \epsilon_\theta(x_{t_{i-1}}, \lambda_{t_{i-1}}) - \sigma_{t_i} B(h_i) \sum_{m=1}^{p-1} \frac{a_m}{r_m} D_m, \tag{I.14}$$

and then corrects $x_{t_i}$ by $x_{t_i}^c$ with the corrector (UniC-p)

$$x_{t_i}^c = \frac{\alpha_{t_i}}{\alpha_{t_{i-1}}} x_{t_{i-1}}^c - \sigma_{t_i}(e^{h_i} - 1) \epsilon_\theta(x_{t_{i-1}}, \lambda_{t_{i-1}}) - \sigma_{t_i} B(h_i) \sum_{m=1}^{p} \frac{a_m}{r_m} D_m. \tag{I.15}$$

Here $x_{t_i}^c$ is a corrected result of $x_{t_i}$, $B(h) = O(h)$ is a function of $h$, $a_m$ is a determined coefficient, and $D_m$ is computed as

$$D_m = \epsilon_\theta(x_{s_m}, \lambda_{s_m}) - \epsilon_\theta(x_{t_{i-1}}, \lambda_{t_{i-1}}). \tag{I.16}$$

One can show that UniP-p has $p$-th order of accuracy, and UniC-p has $(p+1)$-th order of accuracy (see Theorem 3.1 and Corollary 3.2 in Ref. [10]).

The distinct mathematical structures of DPM-solver and UniPC in solving the diffusion ODE lead to different strengths and limitations. Firstly, DPM-solver employs exact $k$-th order derivatives to construct a Taylor approximation of the denoising function $\epsilon_\theta(x_\lambda, \lambda)$, which offers higher accuracy and precision, making it ideal for tasks requiring detailed computations in situations where computational resources do not constitute a major constraint. Secondly, UniPC, in contrast, approximates higher-order derivatives using finite differences of $\epsilon_\theta(x_\lambda, \lambda)$ at various points $(x_{s_m}, \lambda_m)$. This method prioritizes efficiency and ease of implementation, making it well-suited for large-scale or resource-constrained applications. Each approach is tailored to specific use cases, with DPM-solver excelling in precision-critical scenarios and UniPC thriving in contexts demanding computational efficiency. A recent work studied UniPC with optimized time steps to improve image generation performance [12]. The proposed approach introduces optimized time-stepping methods for diffusion sampling, reducing the number of steps while maintaining or improving sample quality exploring both the DPM-solver technique and the UniPC approach. Both methods are evaluated for their ability to accelerate sampling while balancing computational cost and accuracy, paving the way for faster, more practical implementations of DPMs in real-world applications.

## II. EFFICIENT QUANTUM ALGORITHMS FOR DPMS

We now turn to presenting efficient quantum algorithms. In our approach, we build on and substantially generalize and further develop the methodology of Liu et al. [13]. In this work, it has been demonstrated that fault-tolerant quantum computing could provide provably efficient solutions for generic (stochastic) gradient descent algorithms, a cornerstone of machine learning, giving rise to the presumably to date most practically minded application in the field. The key innovation lies in employing the technique developed in Ref. [14], which leverages Carleman linearization. This method transforms a system of non-linear differential equations into an infinite-dimensional system of linear differential equations. By applying this approach to (stochastic) gradient descent algorithms, quantum computers can potentially achieve significant efficiency gains in machine learning tasks.

### A. Quantum Carleman DPM-solver

In what follows, the goal is to develop an efficient quantum algorithm to enhance the ODE solvers discussed above. To achieve this, we begin with the DPM-solver outlined in Eq. (I.13), restated as

$$x_{t_i} = \frac{\alpha_{t_i}}{\alpha_{t_{i-1}}} x_{t_{i-1}} - \alpha_{t_i} \sum_{n=0}^{k-1} \epsilon_\theta^{(n)}(x_{\lambda_{t_{i-1}}}, \lambda_{t_{i-1}}) \int_{\lambda_s}^{\lambda_t} e^{-\lambda} \frac{(\lambda - \lambda_{t_{i-1}})^n}{n!} \mathrm{d}\lambda. \tag{II.1}$$

We outline the logic and structure of the proposed quantum algorithm as follows.

- First, we express $\epsilon_\theta(x_\lambda, \lambda)$ and its higher-order derivatives as a Taylor expansion in terms of $x_\lambda$. From this, we derive a forward iteration of the form $x_{t_i} = \text{poly}(x_{t_{i-1}})$.

- Since this represents a non-linear ODE, we employ *Carleman linearization* to transform it into a linear system, enabling the application of quantum algorithms.

- We define $y_j(\lambda) \approx x_\lambda^j$ for $j = 1, \ldots, N$. As the Carleman linearization transforms a non-linear differential equation into an infinite-dimensional system of linear differential equations, we truncate the system at $N$, with the initial condition $y_j(\lambda_{t_{i-1}}) = x_{\lambda_{t_{i-1}}}^j$.

- By letting $Y = (y_1, \ldots, y_N)$, we derive a linear iteration of the form $Y(\lambda_{t_i}) = AY(\lambda_{t_{i-1}})$ based on Eq. (II.1) with matrix $A$, which is the *quantum Carleman matrix* (QCM) associated with the DPM-solver.

- With the linear iteration equations established, for every time step $\{t_i\}_{i=0}^M$ with $t_0 = T$ and $t_M = 0$, we can construct a system of linear equations and apply *quantum linear system solvers* (QLSS) to compute $(Y(\lambda_{t_0}), \ldots, Y(\lambda_{t_M}))$ and measure the amplitude proportional to $y_1(t_M) = y_1(0) = x(0)$.

The remainder of this subsection will provide a detailed explanation of these steps and the underlying logic.

## 1. Polynomial iteration

We now present the details of the processes described above. First, we rewrite

$$\epsilon_\theta(x_\lambda, \lambda) = \sum_{j=0}^{J} \frac{a_j}{j!} x_\lambda^j \,. \tag{II.2}$$

In other words, we consider $\epsilon_\theta(x_\lambda, \lambda)$ as an order $J$ polynomial. Note that we take into account only values $k \le J$. Then we find the $n$-th derivative to be

$$\epsilon_\theta^{(n)}(x_\lambda, \lambda) = \sum_{j=0}^{J-n} \frac{a_{j+n}}{j!} x_\lambda^j \,. \tag{II.3}$$

We then let $x_\lambda = x_{t_{i-1}}$ and plug this into Eq. (II.1), to get

$$x_{t_i} = \frac{\alpha_{t_i}}{\alpha_{t_{i-1}}} x_{t_{i-1}} - \alpha_{t_i} \sum_{n=0}^{k-1} \left( \sum_{j=0}^{J-n} \frac{a_{j+n}}{j!} x_{t_{i-1}}^j \right) \int_{\lambda_s}^{\lambda_t} e^{-\lambda} \frac{(\lambda - \lambda_{t_{i-1}})^n}{n!} \mathrm{d}\lambda \,. \tag{II.4}$$

We first take care of the integral at the end, following the treatment in Ref. [5]. To this end, first define the function

$$\varphi_k(z) := \int_0^1 e^{(1-\delta)z} \frac{\delta^{k-1}}{(k-1)!} d\delta \,, \quad \varphi_0(z) = e^z \,, \quad \varphi_k(0) = \frac{1}{k!} \,, \tag{II.5}$$

which satisfies the recurrence relation

$$\varphi_{k+1}(z) = \frac{\varphi_k(z) - \varphi_k(0)}{z} \,. \tag{II.6}$$

According to Ref. [5], Appendix B.2, we can rewrite Eq. (II.4) as

$$x_{t_i} = \frac{\alpha_{t_i}}{\alpha_{t_{i-1}}} x_{t_{i-1}} - \sigma_{t_i} \sum_{n=0}^{k-1} h_i^{n+1} \varphi_{n+1}(h_i) \left( \sum_{j=0}^{J-n} \frac{a_{j+n}}{j!} x_{t_{i-1}}^j \right) + \mathcal{O}\left( h_i^{k+1} \right) \,, \tag{II.7}$$

where $h_i = \lambda_{t_i} - \lambda_{t_{i-1}}$. By the recurrence relation Eq. (II.6), we can deduce the general formula for the function $\varphi_{n+1}(h_i)$ as

$$\varphi_{n+1}(h_i) = \frac{e^{h_i} - \sum_{\ell=0}^{n} \frac{h_i^\ell}{\ell!}}{h_i^{n+1}} = \frac{\sum_{\ell=n+1}^{\infty} \frac{h_i^\ell}{\ell!}}{h_i^{n+1}} \,. \tag{II.8}$$

Plugging this form of $\varphi_{n+1}(h_i)$ into Eq. (II.7), we find

$$x_{t_i} = \frac{\alpha_{t_i}}{\alpha_{t_{i-1}}} x_{t_{i-1}} - \sigma_{t_i} \sum_{n=0}^{k-1} \left( \sum_{\ell=n+1}^{k-1} \frac{h_i^\ell}{\ell!} \right) \left( \sum_{j=0}^{J-n} \frac{a_{j+n}}{j!} x_{t_{i-1}}^j \right) \,. \tag{II.9}$$

We then define in a next step

$$f^{(n+1)}(h_i) := \sum_{\ell=n+1}^{k-1} \frac{h_i^\ell}{\ell!} \quad \text{for } n \ge 0 \,. \tag{II.10}$$

As a result, we rewrite Eq. (II.4) in terms of this $f^{(n)}(h_i)$ function as

$$\delta x_{t_i} = \left( \frac{\alpha_{t_i}}{\alpha_{t_{i-1}}} - 1 \right) x_{t_{i-1}} - \sigma_{t_i} \sum_{n=0}^{k-1} \left( \sum_{j=0}^{J-n} \frac{a_{j+n}}{j!} x_{t_{i-1}}^j \right) f^{(n+1)}(h_i) \,. \tag{II.11}$$

Hence, we have obtained a polynomial expression for this difference equation. The next step is to do Carleman linearization for Eq. (II.11).

## 2. Carleman linearization

We here lay out details of the Carleman linearization presented in the main text. We first write the polynomial above in a more concise fashion as

$$\delta(x_{t_i}) = \sum_{j=0}^{J} F_j \, x_{t_i}^j \, , \tag{II.12}$$

where for $F_1$, we find

$$F_1 = \left( \frac{\alpha_{t_i}}{\alpha_{t_{i-1}}} - 1 \right) - \sigma_{t_i} \left( a_j f^{(1)} + a_{j+1} f^{(2)} + \ldots + a_{j+k-1} f^{(k)} \right) \, , \tag{II.13}$$

and, for $j \neq 1$ or $F_0$ and $F_{j \geq 2}$, we get

$$F_{j \neq 1} = \begin{cases} -\frac{\sigma_{t_i}}{j!} \left( a_j f^{(1)} + a_{j+1} f^{(2)} + \ldots + a_{j+k-1} f^{(k)} \right) & \text{for } 0 \leq j \leq J - (k-1) \, , \\ -\frac{\sigma_{t_i}}{j!} \left( a_j f^{(1)} + a_{j+1} f^{(2)} + \ldots + a_{J-1} f^{(J-j)} + a_J f^{(J-j+1)} \right) & \text{for } J - k + 2 \leq j \leq J \, , \end{cases} \tag{II.14}$$

where

$$f^{(n+1)}(h_i) = \sum_{\ell=n+1}^{k-1} \frac{h_i^\ell}{\ell!} \, . \tag{II.15}$$

Note that the functions $f^{(n)}$ are of leading order $\mathcal{O}(h_i^n)$ in $h_i$. Eq. (II.12) presents a $J$-th degree polynomial differential equation, with "initial" condition given by (after $M$ iteration)

$$x^{\otimes 1}(t_M) = x^{\otimes 1}(0) = x(0) \, . \tag{II.16}$$

Recall that $x(t_i) \in \mathbb{R}^D$, we will write explicitly $x^{\otimes q}$ to emphasize the fact that we are dealing with the general vector defined as

$$\begin{aligned}
&F_0 \in \mathbb{R}^{D \times 1} \, , \\
&x(t_i) = (x_1, \ldots, x_D) \in \mathbb{R}^D, F_1 \in \mathbb{R}^{D \times D} \, , \\
&x^{\otimes 2} = \left( x_1^2, x_1 x_2, \ldots, x_1 x_D, x_2 x_1, \ldots, x_D x_{D-1}, x_D^2 \right) \in \mathbb{R}^{D^2}, F_2 \in \mathbb{R}^{D \times D^2} \, , \\
&x^{\otimes 3} = \left( x_1^3, x_1^2 x_2, \ldots, \right) \in \mathbb{R}^{D^3}, F_3 \in \mathbb{R}^{D \times D^3} \, , \\
&\ldots \\
&x^{\otimes J} = \left( x_1^J, x_1^{J-1} x_2, \ldots, \right) \in \mathbb{R}^{D^J}, F_J \in \mathbb{R}^{D \times D^J} \, .
\end{aligned} \tag{II.17}$$

It is important to note the difference between this model compared to the analytical structure presented in Ref. [13]. In Ref. [13], the gradient descent algorithm have initial condition $u(0) = u_{\text{in}}$, whereas in the case of DPMs, the process moves forward by adding noise and reverses to eliminate noise during the backward pass. Consequently, our initial condition corresponds to the final iteration in this context. Furthermore, one can define matrix elements as

$$\begin{aligned}
&A_{j+J-1}^j = F_J \otimes I^{\otimes j-1} + I \otimes F_J \otimes I^{\otimes j-2} + \cdots + I^{\otimes j-1} \otimes F_J \, , \\
&\ldots \\
&A_{j+1}^j = F_2 \otimes I^{\otimes j-1} + I \otimes F_2 \otimes I^{\otimes j-2} + \cdots + I^{\otimes j-1} \otimes F_2 \, , \\
&A_j^j = F_1 \otimes I^{\otimes j-1} + I \otimes F_1 \otimes I^{\otimes j-2} + \cdots + I^{\otimes j-1} \otimes F_1 \, , \\
&A_{j-1}^j = F_0 \otimes I^{\otimes j-1} + I \otimes F_0 \otimes I^{\otimes j-2} + \cdots + I^{\otimes j-1} \otimes F_0 \, .
\end{aligned} \tag{II.18}$$

Now we are in the place the formulate the quantum Carleman linearization for the DPM-solver as

$$\delta \begin{pmatrix} x^{\otimes 1} \\ x^{\otimes 2} \\ x^{\otimes 3} \\ \vdots \\ x^{\otimes(N-1)} \\ x^{\otimes N} \\ \vdots \end{pmatrix} = \begin{pmatrix} A_1^1 & A_2^1 & \cdots & A_J^1 & & & \\ A_1^2 & A_2^2 & \cdots & A_J^2 & A_{J+1}^2 & & \\ & A_2^3 & A_3^3 & A_4^3 & \cdots & & \\ & & \ddots & \ddots & \ddots & & \\ & & & A_{N-2}^{N-1} & A_{N-1}^{N-1} & A_N^{N-1} & \cdots \\ & & & & A_{N-1}^N & A_N^N & \cdots \\ & & & & & \ddots & \ddots \end{pmatrix} \begin{pmatrix} x^{\otimes 1} \\ x^{\otimes 2} \\ x^{\otimes 3} \\ \vdots \\ x^{\otimes(N-1)} \\ x^{\otimes N} \\ \vdots \end{pmatrix} + \begin{pmatrix} F_0 \\ 0 \\ 0 \\ \vdots \\ 0 \\ 0 \\ \vdots \end{pmatrix} \, . \tag{II.19}$$

Note that we can express the differential equation Eq. (II.12) in terms of the matrix elements defined above,

$$\delta(x^j(t_i)) = \sum_{k=0}^{J} G_{j,k} \, x^{k+j-1}(t_i) \, ,$$ (II.20)

where

$$G_{j,k} = \text{vec}\left(\sum_{\ell=0}^{j-1} I^{\otimes(\ell)} \otimes F_k \otimes I^{\otimes(j-1-\ell)}\right) = A_{k+j-1}^{j} \, .$$ (II.21)

Based on these expressions, we can construct the linear system truncated with the parameter $N$. We first define $\hat{y}_j(\lambda) \approx x^{\otimes j}(\lambda)$ for $j = 1, \dots, N$. This means that we truncate the polynomial at order $N$. The differential equation for $\hat{y}_j(\lambda)$ is

$$\delta\hat{y} = A\hat{y} + b, \quad y_1(t_M) = y_1(0) = x_{\text{in}}(0) \, ,$$ (II.22)

which in turn is

$$\delta \begin{pmatrix} \hat{y}_1 \\ \hat{y}_2 \\ \hat{y}_3 \\ \vdots \\ \hat{y}_{N-1} \\ \hat{y}_N \end{pmatrix} = \begin{pmatrix} A_1^1 & A_2^1 & \cdots & A_J^1 & & \\ A_1^2 & A_2^2 & \cdots & A_J^2 & A_{J+1}^2 & \\ & A_2^3 & A_3^3 & A_4^3 & & \\ & & \ddots & \ddots & \ddots & \\ & & & A_{N-2}^{N-1} & A_{N-1}^{N-1} & A_N^{N-1} \\ & & & & A_{N-1}^{N} & A_N^{N} \end{pmatrix} \begin{pmatrix} \hat{y}_1 \\ \hat{y}_2 \\ \hat{y}_3 \\ \vdots \\ \hat{y}_{N-1} \\ \hat{y}_N \end{pmatrix} + \begin{pmatrix} F_0 \\ 0 \\ 0 \\ \vdots \\ 0 \\ 0 \end{pmatrix} \, .$$ (II.23)

We denote the big vector in both sides of Eq. (II.23) as

$$\hat{Y} = (\hat{y}_1, \hat{y}_2, \dots, \hat{y}_N) \, .$$ (II.24)

At this point, we can express the entire iteration process as a single linear differential equation of the form

$$\delta\hat{Y} = \hat{Y}(t_i) - \hat{Y}(t_{i-1}) = A\hat{Y}(t_{i-1}) + b \, , \quad \hat{y}_1(t_M) = \hat{y}_1(0) = x_{\text{in}}(0)$$ (II.25)

of this $\hat{Y}(t_i)$ vector, for all time steps $\{t_i\}_{i=0}^{M}$ with $t_0 = T$ and $t_M = 0$. Now we have a linear system at hand, we can run the state-of-the-art *quantum linear system solver* (QLSS) programs to solve it efficiently.

### 3. Quantum ODE solver

The whole iteration scheme in Eq. (II.25) can be captured in a vector equation in the form $M\boldsymbol{Y} = \boldsymbol{b}$, which combines $\hat{Y}(t_i)$ at all time steps to form a single big vector $\boldsymbol{Y} = (\hat{Y}(t_0), \hat{Y}(t_1), \dots, \hat{Y}(t_M))$, in the form of

$$\begin{pmatrix} -(I+A) & I & & & \\ & \ddots & \ddots & & \\ & & -(I+A) & I & \\ & & & -(I+A) & I \\ & & & & I \end{pmatrix} \begin{pmatrix} \hat{Y}(t_0) \\ \hat{Y}(t_1) \\ \vdots \\ \hat{Y}(t_{M-1}) \\ \hat{Y}(t_M) \end{pmatrix} = \begin{pmatrix} b \\ b \\ \vdots \\ b \\ \hat{X}(0) \end{pmatrix} \, .$$ (II.26)

The first observation that we can make is about the sparsity of the big matrix $M$ in Eq. (II.26). First, we take a look at the Carleman matrix $A$ in Eq. (II.23). Based on the form of the Carleman matrix in Eq. (II.23), both the row and column sparsity are less than or equal to $J$ the order of our polynomial expansion. We denote the sparsity of $A$ as $s_A$, so $s_A \leq J$. The matrix $M$ for our linear system: $M\boldsymbol{Y} = \boldsymbol{b}$ then have the sparsity $s_M$ given by

$$s_M \leq J + 1 \, .$$ (II.27)

The most natural way for solving system like Eq. (II.26) is to implement the *quantum linear system solver* (QLSS), which has a well-established history of development. The first quantum algorithm for solving sparse systems of linear equations was the *Harrow-Hassidim-Lloyd* (HHL) algorithm, proposed in 2008 [15]. The authors have demonstrated that $\mathcal{O}\left(\text{poly}\left(\kappa, \frac{1}{\epsilon}\right)\right)$ queries

are sufficient to produce the solution state with accuracy $\epsilon > 0$ [15]. Here, $\kappa > 0$ represents the spectral condition number of the target linear system. In our case, $\kappa$ is defined as

$$\kappa := \|M\| \, \|M^{-1}\| \, , \tag{II.28}$$

where $\|M\|$ is the operator norm of $M$. The scaling in $\kappa$ has later been refined to nearly linear using Ambainis' *variable time amplitude amplification* (VTAA) algorithm [16]. This improvement was followed by an exponential enhancement in the error scaling to $\mathcal{O}\left(\mathrm{polylog}\frac{1}{\epsilon}\right)$ achieved by employing a linear combination of quantum walks within VTAA [17].

Further advancements have been made through alternative approaches based on adiabatic evolution and notions of eigenstate filtering [18–20], which has resulted in algorithms achieving a similar scaling in $\kappa$ and $\epsilon$. This progression culminated in the development of an algorithm with a scaling of $\mathcal{O}\left(\kappa \log \frac{1}{\epsilon}\right)$ and constant success probability [21], which has been claimed to be optimal. More recently, a simpler quantum algorithm achieving the same $\mathcal{O}\left(\kappa \log \frac{1}{\epsilon}\right)$ scaling has been introduced. This algorithm utilizes kernel reflection on an augmented linear system [22]. In QLSP, two distinct query oracles play a role [23]:

- *Matrix oracle* ($\mathcal{O}_M$): This oracle block encodes the matrix $M$ that needs to be inverted.

- *State preparation oracle* ($\mathcal{O}_{\boldsymbol{b}}$): This oracle prepares the initial state vector $|\boldsymbol{b}\rangle$ from a standard reference state, typically the computational basis state vector $|0\rangle$, without loss of generality.

The query complexities referenced earlier represent the worst-case combined cost, treating both oracles $\mathcal{O}_M$ and $\mathcal{O}_{\boldsymbol{b}}$ as having equal weight in the analysis. However, in practice, the cost associated with $\mathcal{O}_{\boldsymbol{b}}$ can often be significantly lower than that of $\mathcal{O}_M$. This distinction arises because preparing the initial state vector $|\boldsymbol{b}\rangle$ may involve simpler operations compared to the block encoding of the matrix $M$, see Ref. [23]. We will now elaborate on why and how this difference manifests itself.

The process of constructing a block encoding of $M^{-1}/(2\,\|M^{-1}\|)$ relies on the *quantum singular value transformation* (QSVT). This method achieves the desired block encoding with precision $\epsilon$ by making $\mathcal{O}\left(\kappa \log \frac{1}{\epsilon}\right)$ queries to $\mathcal{O}_M$, as detailed in Corollary 69 of Ref. [24]. Once the block encoding is constructed, applying it to the initial state vector $|\boldsymbol{b}\rangle$ produces the unnormalized state vector

$$\frac{M^{-1}|\boldsymbol{b}\rangle}{2\,\|M^{-1}\|}, \tag{II.29}$$

with accuracy $\epsilon$ and success probability $p_{\mathrm{succ}}/4$, where

$$p_{\mathrm{succ}} = \frac{\|A^{-1}|b\rangle\|^2}{\|A^{-1}\|^2}, \quad 1 \le \frac{1}{\sqrt{p_{\mathrm{succ}}}} = \frac{\|A^{-1}\|}{\|A^{-1}|b\rangle\|} \le \kappa \, . \tag{II.30}$$

Importantly, this step requires just one query to the state preparation oracle $\mathcal{O}_{\boldsymbol{b}}$. The probability can be amplified close to unity using amplitude amplification, requiring $\mathcal{O}\left(1/\sqrt{p_{\mathrm{succ}}}\right)$ rounds of. However, this amplification process also increases the error in the solution state proportionally. To achieve an accuracy of $\epsilon$ in the normalized solution state vector $\frac{A^{-1}|b\rangle}{\|A^{-1}|b\rangle\|}$, the total number of queries to the oracles must scale as

$$\mathcal{O}\left(\frac{1}{\sqrt{p_{\mathrm{succ}}}}\mathbf{Cost}\left(\mathcal{O}_{\boldsymbol{b}}\right) + \frac{\kappa}{\sqrt{p_{\mathrm{succ}}}}\log\left(\frac{1}{\sqrt{p_{\mathrm{succ}}}\epsilon}\right)\mathbf{Cost}\left(\mathcal{O}_M\right)\right) \, . \tag{II.31}$$

Thus, the number of queries to $\mathcal{O}_{\boldsymbol{b}}$ achieves a strictly linear scaling with $\frac{1}{\sqrt{p_{\mathrm{succ}}}}$, surpassing nearly all prior results mentioned earlier. This linear scaling can often be significantly smaller than $\kappa$ in practical scenarios. Notably, this scaling was already achieved by the original HHL algorithm introduced in Ref. [15], as later observed and elaborated upon in Ref. [25].

Thus, to conclude, the number of queries to $\mathcal{O}_{\boldsymbol{b}}$ attains a strictly linear scaling with $\frac{1}{\sqrt{p_{\mathrm{succ}}}}$ surpassing almost all previous results cited above and can be much smaller than $\kappa$ in practice. In fact this scaling has already been achieved by the algorithm in the original HHL algorithm in Ref. [15] as observed in Ref. [25]. Methods leveraging the adiabatic theorem and the kernel reflection technique, mentioned before [21, 22], exhibit a scaling of

$$\mathcal{O}\left(\kappa \log\left(\frac{1}{\epsilon}\right)\mathbf{Cost}\left(\mathcal{O}_{\boldsymbol{b}}\right) + \kappa \log\left(\frac{1}{\epsilon}\right)\mathbf{Cost}\left(\mathcal{O}_M\right)\right) \, . \tag{II.32}$$

However, in these methods, the complexity of initial state preparation $\mathcal{O}_{\boldsymbol{b}}$ can become significantly worse compared to earlier approaches.

To date, the only known method achieving near-optimal complexity for both $\mathcal{O}_{\boldsymbol{b}}$ and $\mathcal{O}_M$ is based on VTAA. The state-of-the-art VTAA algorithm requires $\mathcal{O}\left(p_{\mathrm{succ}}^{-1/2} \log(\kappa)\right)$ queries to $\mathcal{O}_{\boldsymbol{b}}$, assuming the intermediate success probabilities of the

algorithm can be estimated to a constant multiplicative accuracy [26, 27]. If this prior knowledge about intermediate success probabilities is unavailable, the query complexity worsens to $\mathcal{O}\left(p_{\text{succ}}{}^{-1/2}\log^3(\kappa)\log\log(\kappa)\right)$, While this scenario increases complexity, it introduces only a one-time cost when VTAA is compiled into a quantum circuit. However, it does make the algorithm significantly more intricate, highlighting the importance of accurate success probability estimation for optimal performance.

Recently, Low et al. in Ref. [23] proposed a quantum linear system algorithm that emphasizes improving the query complexity of initial state preparation. By leveraging the block preconditioning technique, their algorithm achieves a query complexity of $\mathcal{O}\left(\kappa\log\left(\frac{1}{\epsilon}\right)\right)$ for both $\mathcal{O}_M$ (matrix block encoding) and $\mathcal{O}_b$ (initial state preparation). This method, along with the kernel-reflection approach [22] and the discrete adiabatic theorem-based approach [21], represents the state-of-the-art in query complexity for solving quantum linear systems, as stated in Eq. (II.32) offering significant advancements in efficiency and practicality.

The linear combination of Hamiltonian simulation (LCHS) strategy [28, 29] provides an alternative approach to solving Eq. (II.25) by representing the solution as a linear combination of Hamiltonian simulations. Consider a system of the form

$$\partial_t u(t) = -A(t)u(t) + b(t), \quad u(0) = u_0,$$ (II.33)

which matches the structure of Eq. (II.26). The LCHS method is based on the remarkable identity that expresses the solution in terms of a weighted integral of Hamiltonian simulations. Specifically, by decomposing the matrix $A(t)$ into its Hermitian and anti-Hermitian components,

$$A(t) = L(t) + iH(t),$$ (II.34)

where $L(t) \geq 0$ for all $t \in [0, T]$, Theorem 1 in Ref. [28] provides the identity

$$\mathbf{T}\, e^{-\int_0^t A(s)\mathrm{d}s} = \int_{\mathbb{R}} \frac{1}{\pi(1+k^2)} \mathbf{T}\, e^{-i\int_0^t (H(s)+kL(s))\mathrm{d}s}\, \mathrm{d}k,$$ (II.35)

where $\mathbf{T}$ denotes the time-ordering operator. When the source term $b(t)$ is present, Duhamel's principle (also known as variation of constants) can be applied to express the solution as

$$
\begin{aligned}
u(t) = {} & \int_{\mathbb{R}} \frac{1}{\pi(1+k^2)} \mathbf{T}\, e^{-i\int_0^t (H(s)+kL(s))\mathrm{d}s} u_0\, \mathrm{d}k \\
& + \int_0^t \int_{\mathbb{R}} \frac{1}{\pi(1+k^2)} \mathbf{T}\, e^{-i\int_s^t (H(s')+kL(s'))\mathrm{d}s'} b(s)\, \mathrm{d}k\, \mathrm{d}s.
\end{aligned}
$$ (II.36)

The LCHS approach can be efficiently implemented by combining linear combinations of unitaries (LCU) [30] with existing Hamiltonian simulation algorithms. In practice, the integral in Eq. (II.35) is truncated to a finite interval $[-K, K]$ and discretized using the trapezoidal rule with $M+1$ grid points.

Compared to the *quantum linear system solver* (QLSS) approach, the LCHS method provides significant advantages. It directly implements the time evolution operator without relying on QLSS, thereby reducing the number of state preparation oracle calls. For instance, when $b(t) = 0$, the LCHS algorithm requires

$$\mathcal{O}\left(\frac{\|u_0\|}{\|u(T)\|}\right)$$ (II.37)

queries to the state preparation oracle, whereas the state-of-the-art QLSA-based quantum Dyson series method [31] requires

$$\mathcal{O}\left(\frac{\|u_0\|}{\|u(T)\|}\|A\|T\log(1/\epsilon)\right)$$ (II.38)

many queries. The LCHS method thus eliminates explicit dependence on parameters such as $\|A\|$, $T$, and $\epsilon$, achieving optimal scaling and offering a more efficient alternative to traditional QLSS-based techniques.

### B. Quantum Carleman UniPC

In this subsection, we aim to design a quantum Carleman algorithm tailored for the alternative technique to solve the Diffusion ODE: UniPC. This includes both the predictor (UniP-p) and the corrector (UniC-p), as defined respectively in Eq. (I.14) and Eq. (I.15), restated as

$$x_{s_p} = \frac{\alpha_{s_p}}{\alpha_{s_0}} x_{s_0} - \sigma_{s_p}(e^{h_i} - 1)\epsilon_\theta(x_{s_0}, \lambda_{s_0}) - \sigma_{s_p} B(h_i) \sum_{m=1}^{p-1} \frac{a_m}{r_m} D_m,$$ (II.39)

and as

$$x_{s_p}^c = \frac{\alpha_{s_p}}{\alpha_{s_0}} x_{s_0}^c - \sigma_{s_p}(e^{h_i} - 1)\epsilon_\theta(x_{s_0}, \lambda_{s_0}) - \sigma_{s_p} B(h_i) \sum_{m=1}^{p} \frac{a_m}{r_m} D_m, \tag{II.40}$$

with finite difference $D_m = \epsilon_\theta(x_{s_m}, \lambda_{s_m}) - \epsilon_\theta(x_{s_0}, \lambda_{s_0})$ and interpolation time steps $s_m = h_i r_m + \lambda_{t_{i-1}}$ with $m = 1, 2, \ldots, p$. Note that $s_0 = \lambda_{t_{i-1}}$, and $s_p = \lambda_{t_i}$.

Again, we first outline the logic and structure of the proposed quantum algorithm as follows.

- To start with, as what we did for the DPM-solver algorithm, we rewrite the denoising function $\epsilon_\theta(x_\lambda, \lambda)$ as a Taylor expansion in terms of $x_\lambda$ (up to order $J$ expansion):

$$\epsilon_\theta(x_\lambda, \lambda) = \sum_{j=0}^{J} \frac{a_j}{j!} x_\lambda^j, \tag{II.41}$$

where $a_j$ contains $\lambda$.

- We are then able to express $\epsilon_\theta(x_\lambda, \lambda)$ at different points as a polynomial of $x_\lambda$, and derive a multi-step forward iteration for UniP: $x_{s_p} = \mathrm{poly}(x_{s_0}) + \mathrm{poly}(x_{s_1}) + \cdots + \mathrm{poly}(x_{s_{p-1}})$; and for UniC: $x_{s_p}^c = x_{s_0}^c + \mathrm{poly}(x_{s_0}) + \mathrm{poly}(x_{s_1}) + \cdots + \mathrm{poly}(x_{s_p})$.

- We employ *Carleman linearization* in this step and truncate the system with the given parameter $N$. We denote $y_j(\lambda) \approx x_\lambda^j(\lambda)$, $z_j(\lambda) \approx (x^c)_\lambda^j(\lambda)$ for $j = 1, \ldots, N$, with the initial condition $y_1(s_0) = x_\lambda(s_0)$, $z_1(s_0) = (x^c)_\lambda(s_0)$.

- Let $Y = (y_1, \ldots, y_N)$, and $Z = (z_1, \ldots, z_N)$, we utilize Eq. (II.39) and Eq. (II.40) to derive a multi-step linear iteration (predictor) $Y(s_p) = A_0 Y(s_0) + \cdots + A_{p-1} Y(s_{p-1})$ with matrices $A_0, \ldots, A_{p-1}$; and another linear iteration (corrector) $Z(\lambda_{t_i}) = BZ(\lambda_{t_{i-1}}) + A_0 Y(s_0) + \cdots + A_p Y(s_p)$ with matrices $B, A_0, \ldots, A_p$.

- For all time steps $\{t_i\}_{i=0}^{M}$ with $t_0 = T$ and $t_M = 0$, we construct a system of linear equations and implement a *quantum linear system solver* (QLSS) to obtain $(Y(\lambda_{t_0}), \ldots, Y(\lambda_{t_M}))$ and $(Z(\lambda_{t_0}), \ldots, Z(\lambda_{t_M}))$, and measure the amplitude proportion to (the corrected) $z_1(t_M) = z_1(0) = x^c(0)$.

We emphasize that the Carleman linearization structure for UniPC introduced here represents a novel Carleman predictor-corrector numerical scheme. This approach differs from both the DPM-solver framework described in the previous section and the structure presented in Ref. [13]. A defining feature of the quantum Carleman algorithm for UniPC is the generation of a unique *quantum Carleman matrix* (QCM) for each interpolation time step $s_m$.

### 1. Polynomial iteration for UniP-p

To avoid being messy and confusing, we first focus on the UniP part of the algorithm. We will derive a quantum algorithm for the predictor and then we will make use of this result to produce the corrector algorithm for UniC. Again we plug in the polynomial expression Eq. (II.41) into Eq. (II.39), to get

$$x_{s_p} = \frac{\alpha_{s_p}}{\alpha_{s_0}} x_{s_0} - \sigma_{s_p}(e^{h_i} - 1)\left(\sum_{j=0}^{J} \frac{a_j}{j!} x_{s_0}^j\right) - \sigma_{s_p} B(h_i) \sum_{m=1}^{p-1} \frac{a_m}{r_m}\left(\sum_{j=0}^{J} \frac{a_j}{j!}\left(x_{s_m}^j - x_{s_0}^j\right)\right). \tag{II.42}$$

It is not hard to see that this Eq. (II.42) gives a difference equation in the form

$$\begin{aligned}
\delta x &= x_{s_p} - x_{s_0} = \mathrm{poly}(x_{s_0}) + \mathrm{poly}(x_{s_1}) + \ldots + \mathrm{poly}(x_{s_{p-1}}) \\
&= \sum_{j=0}^{J} F_j^{(0)}(\lambda)\, x_{s_0}^j + \sum_{j=0}^{J} F_j^{(1)}(\lambda)\, x_{s_1}^j + \ldots + \sum_{j=0}^{J} F_j^{(p-1)}(\lambda)\, x_{s_{p-1}}^j.
\end{aligned} \tag{II.43}$$

The coefficients of the polynomials in Eq. (II.43) are gathered below. We first find the coefficients $F_j^{(0)}$ of $\mathrm{poly}(x_{s_0})$ to be given by

$$\begin{aligned}
F_1^{(0)} &= \frac{\alpha_{s_p}}{\alpha_{s_0}} - 1 - \sigma_{s_p} a_1\left(e^{h_i} - 1 - B(h_i) \sum_{m=1}^{p-1} \frac{a_m}{r_m}\right), \\
F_j^{(0)} &= -\frac{\sigma_{s_p} a_j}{j!}\left(e^{h_i} - 1 - B(h_i) \sum_{m=1}^{p-1} \frac{a_m}{r_m}\right), \quad \text{for } 0 \le (j \ne 1) \le J.
\end{aligned} \tag{II.44}$$

Furthermore, we find the coefficients $F_j^{(m)}$ of $\text{poly}(x_{s_m})$ to be

$$F_j^{(m)} = -\frac{\sigma_{s_p} B(h_i) a_j a_m}{r_m(j!)} , \quad \text{for } 0 \leq j \leq J \text{ and } m = 1, 2, \ldots, p-1 . \tag{II.45}$$

### 2. Polynomial iteration for UniC-p

We here do the exact same thing for the corrector. In order to do so, we use Eq. (II.40) and obtain the results

$$\delta x^c = \left(\frac{\alpha_{s_p}}{\alpha_{s_0}} - 1\right) x_{s_0}^c - \sigma_{s_p}(e^{h_i} - 1)\left(\sum_{j=0}^{J} \frac{a_j}{j!} x_{s_0}^j\right) - \sigma_{s_p} B(h_i) \sum_{m=1}^{p} \frac{a_m}{r_m}\left(\sum_{j=0}^{J} \frac{a_j}{j!}\left(x_{s_m}^j - x_{s_0}^j\right)\right). \tag{II.46}$$

Again we regroup these terms and obtain the polynomial expression

$$\begin{aligned}
\delta x^c &= \left(\frac{\alpha_{s_p}}{\alpha_{s_0}} - 1\right) x_{s_0}^c + \text{poly}(x_{s_0}) + \text{poly}(x_{s_1}) + \ldots + \text{poly}(x_{s_p}) , \\
&= \sum_{j=0}^{J} \tilde{F}_j \left(x_{s_0}^c\right)^j + \sum_{j=0}^{J} F_j^{(0)_c} x_{s_0}^j + \sum_{j=0}^{J} F_j^{(1)_c} x_{s_1}^j + \ldots + \sum_{j=0}^{J} F_j^{(p)_c} x_{s_p}^j .
\end{aligned} \tag{II.47}$$

The coefficient for $x^c$ is given by

$$\tilde{F}_j = \left(\frac{\alpha_{s_p}}{\alpha_{s_0}} - 1\right) \delta_{j,1} . \tag{II.48}$$

The coefficients for $x$ are

$$F_j^{(0)_c} = -\frac{\sigma_{s_p} a_j}{\alpha_{s_p}(j!)} \left(e^{h_i} - 1 - B(h_i) \sum_{m=1}^{p} \frac{a_m}{r_m}\right) , \quad \text{for } 0 \leq j \leq J , \tag{II.49}$$

and

$$F_j^{(m)_c} = -\frac{\sigma_{s_p} B(h_i) a_j a_m}{\alpha_{s_p} r_m(j!)} , \quad \text{for } 0 \leq j \leq J \text{ and } m = 1, 2, \ldots, p . \tag{II.50}$$

### 3. Carleman linearization (UniP-p)

Similar to what we have done in the previous section for the Carleman linearization of DPM-Solver case, we treat all terms in the polynomial as independent terms. Here, we have much more terms compared to the case before, but we can write each term in a matrix form and put everything together to make up a block diagonal matrix. We first define

$$y_j(t_{i+1}) \approx x_{t_{i+1}}^j \Longrightarrow y_j(s_m) \approx x_{s_m}^j . \tag{II.51}$$

For every time step, we have $p$ vectors given by the interpolation time steps $s_m$: $Y^{(m)}(t_{i-1})$ (for $m = 0, 1, \ldots, p-1$) for which we truncate again at order $N$

$$Y^{(m)}(t_{i-1}) = \begin{pmatrix} y_1(s_m) \\ y_2(s_m) \\ \vdots \\ y_N(s_m) \end{pmatrix} . \tag{II.52}$$

Note that since if $s_0 = t_{i-1}$, then $s_p = t_i$, then we have $Y^{(p)}(t_{i-1}) = Y^{(0)}(t_i) \equiv Y(t_i)$. And so given the difference equation we had before, we can deduce that

$$\delta\left(y_j\right) = \sum_{k=0}^{J} \left(G^{(0)}\right)_k^j y_{k+j-1}(s_0) + \sum_{k=0}^{J} \left(G^{(1)}\right)_k^j y_{k+j-1}(s_1) + \ldots + \sum_{k=0}^{J} \left(G^{(p-1)}\right)_k^j y_{k+j-1}(s_{p-1}) . \tag{II.53}$$

With the same argument as last section, we can express all the non-zero matrix elements, for all the matrices $A^{(m)}$ at all interpolation time steps, gathered as

$$\left(G^{(m)}\right)_k^j = \left(A^{(m)}\right)_{j+k-1}^j = \text{vec}\left[\sum_{\ell=0}^{j-1} I^{\otimes \ell} \otimes F_k^{(m)} \otimes I^{\otimes (j-1-\ell)}\right] \quad \text{for } 0 \leq k \leq J \text{ and } 0 \leq m \leq p-1 . \tag{II.54}$$

For the value of the coefficients $F_k^{(m)}$ refer to Eq. (II.44) and Eq. (II.45). There is also a constant vector

$$b^{(m)} = \begin{pmatrix} F_0^{(m)} \\ 0 \\ \vdots \\ 0 \end{pmatrix} \quad \text{for } m = 0, 1, \ldots, p-1 . \tag{II.55}$$

We gather the above construction and write it as

$$\delta Y = Y(t_i) - Y(t_{i-1}) = A^{(0)} Y^{(0)}(t_{i-1}) + b^{(0)} \\ + A^{(1)} Y^{(1)}(t_{i-1}) + b^{(1)} + \ldots + A^{(p-1)} Y^{(p-1)}(t_{i-1}) + b^{(p-1)} , \tag{II.56}$$

where we explicitly see the difference between the structure of Carleman linearization for the predictor and corrector model in Eq. (II.56) and the structure for DPM-solver presented in Eq. (II.25). The iteration for predictor model also include all the interpolation time steps each associated with a Carleman matrix, whereas the Carleman linearization structure for DPM-solver takes the usual form as Ref. [13].

### 4. Carleman linearization (UniC-p)

In the same spirit, we can proceed with Eq. (II.47). The difference here is the upper bound of the summation over $m$ and an extra linear term of $x_{s_0}^c$, but the structure of the linearization is the same. We define $z_j(t_i) \approx \left(x_{t_i}^c\right)^j$ and $z_j(t_{i+1}) \approx \left(x_{t_{i+1}}^c\right)^j$. Hence from the differential equation structure presented in Eq. (II.47), and the coefficients defined in Eq. (II.48), Eq. (II.49), and Eq. (II.50), we have

$$\delta\left(z_j\right) = \sum_{k=0}^{J} \tilde{G}_k^j \, z_{k+j-1}(s_0) \\ + \sum_{k=0}^{J} \left(G^{(0)_c}\right)_k^j y_{k+j-1}(s_0) + \sum_{k=0}^{J} \left(G^{(1)_c}\right)_k^j y_{k+j-1}(s_1) + \ldots + \sum_{k=0}^{J} \left(G^{(p)_c}\right)_k^j y_{k+j-1}(s_p) , \tag{II.57}$$

where the non-zero matrix elements for Carleman matrix for the corrector $B$ and the predictor $A^{(m)_c}$ are

$$\tilde{G}_k^j = B_{j+k-1}^j = \text{vec}\left[\sum_{\ell=0}^{j-1} I^{\otimes \ell} \otimes \tilde{F}_k \otimes I^{\otimes (j-1-\ell)}\right] , \\ \left(G^{(m)_c}\right)_k^j = \left(A^{(m)_c}\right)_{j+k-1}^j = \text{vec}\left[\sum_{\ell=0}^{j-1} I^{\otimes \ell} \otimes F_k^{(m)_c} \otimes I^{\otimes (j-1-\ell)}\right] , \quad \text{for } m = 0, 1, \ldots, p . \tag{II.58}$$

Again, we define the constant vector $b^{(m)_c}$ for $m = 0, 1, \ldots, p$ and the total vector $Z$ to be

$$b^{(m)_c} = \begin{pmatrix} F_0^{(m)_c} \\ 0 \\ \vdots \\ 0 \end{pmatrix} , \quad Z = \begin{pmatrix} z \\ z^2 \\ \vdots \\ z^N \end{pmatrix} . \tag{II.59}$$

As a result, the matrix equation will take the form

$$\delta Z = Z(t_i) - Z(t_{i-1}) = BZ(t_{i-1}) \\ + A^{(0)_c} Y^{(0)}(t_{i-1}) + b^{(0)_c} + A^{(1)_c} Y^{(1)}(t_{i-1}) + b^{(1)_c} + \ldots + A^{(p)_c} Y^{(p)}(t_{i-1}) + b^{(p)_c} . \tag{II.60}$$

## 5. Quantum algorithm for UniP-p

With Eq. (II.56) defining the iteration scheme for the predictor and Eq. (II.56) for the corrector, we are now positioned to utilize the QLSS in a manner similar to the quantum algorithm for the DPM-solver. The first step is to construct a vector equation that captures the iteration at all time steps forming a huge matrix. For the predictor, we first write all the vectors at different interpolation time steps together forming a big vector $\boldsymbol{Y}(t_i)$ which represents one iteration time step by defining

$$\boldsymbol{Y}(t_i) = \begin{pmatrix} Y^{(0)}(t_i) \\ Y^{(1)}(t_i) \\ \vdots \\ Y^{(p-1)}(t_i) \end{pmatrix} . \tag{II.61}$$

Writing all iteration time steps $\{t_i\}_{i=0}^M$ (for $t_0 = T$ and $t_M = 0$) together, we define one huge vector $\mathbf{Y}$ given by

$$\mathbf{Y} = \begin{pmatrix} \boldsymbol{Y}(t_0) \\ \boldsymbol{Y}(t_1) \\ \vdots \\ \boldsymbol{Y}(t_M) \end{pmatrix} . \tag{II.62}$$

Now we can write Eq. (II.56) in the form $\mathbf{MY} = \mathbf{b}$, where

$$\mathbf{M} = \begin{pmatrix} -(I+A^{(0)}) & -A^{(1)} & \cdots & -A^{(p-1)} & & I & & & & \\ 0 & I & 0 & \cdots & & & & & & \\ & & \ddots & & & & & & & \\ 0 & \cdots & 0 & I & & 0 & \cdots & & & \\ 0 & 0 & \cdots & 0 & -(I+A^{(0)}) & -A^{(1)} & \cdots & -A^{(p-1)} & I & \\ & & & & I & 0 & \cdots & & & \\ & & & & & \ddots & & & & \\ & & & & & & \ddots & & \ddots & \\ & & & & & & & & & I \end{pmatrix} \tag{II.63}$$

and

$$\mathbf{b} = \begin{pmatrix} b^{(0)}(t_0) + b^{(1)}(t_0) + \cdots + b^{(p-1)}(t_0) \\ Y^{(1)}(t_0) \\ \vdots \\ Y^{(p-1)}(t_0) \\ b^{(0)}(t_1) + b^{(1)}(t_1) + \cdots + b^{(p-1)}(t_1) \\ Y^{(1)}(t_1) \\ \vdots \\ \vdots \\ b^{(0)}(t_M) + b^{(1)}(t_M) + \cdots + b^{(p-1)}(t_M) \\ Y^{(1)}(t_M) \\ \vdots \\ Y^{(p-1)}(t_M) \\ X(0) \end{pmatrix} \tag{II.64}$$

where the initial condition for $y(t_M) = y_{\text{in}} = x(0)$ , so we define

$$X(0) = \begin{pmatrix} x(0) \\ x^2(0) \\ \vdots \\ x^N(0) \end{pmatrix} . \tag{II.65}$$

This concluded the part for the predictor. Note that we are not measuring the amplitude for the predictor since the corrector provides a better result as remarked in section I C. So we move on to the discussion on the Carleman linearization for the corrector part of the algorithm.

### 6. Quantum algorithm for UniC-p

For the corrector algorithm, we follow almost the same construction as the predictor model. Again, first, we define a big vector

$$
\boldsymbol{Z}(t_i) = \begin{pmatrix} Z(t_i) \\ Y^{(0)}(t_i) \\ Y^{(1)}(t_i) \\ \vdots \\ Y^{(p)}(t_i) \end{pmatrix}
\tag{II.66}
$$

for one iteration time step. Next, we would like to write all $\boldsymbol{Z}(t_i)$ at all time steps together, forming a huge vector

$$
\mathbf{Z} = \begin{pmatrix} \boldsymbol{Z}(t_0) \\ \boldsymbol{Z}(t_1) \\ \vdots \\ \boldsymbol{Z}(t_M) \end{pmatrix} .
\tag{II.67}
$$

Writing Eq. (II.60) in the form $\mathbf{M}^c \mathbf{Z} = \mathbf{b}^c$, we have the definitions

$$
\mathbf{M}^c = \begin{pmatrix}
-(I+B) & -A^{(0)_c} & -A^{(0)_c} & \cdots & -A^{(p)_c} & I & & & & \\
0 & I & 0 & 0 & \cdots & & & & & \\
0 & 0 & I & 0 & \cdots & & & & & \\
& & & \ddots & & & & & & \\
0 & 0 & \cdots & 0 & I & 0 & \cdots & & & \\
0 & 0 & \cdots & 0 & 0 & -(I+B) & -A^{(1)_c} & \cdots & -A^{(p)_c} & I \\
& & & & & I & 0 & \cdots & & \\
& & & & & & \ddots & \ddots & & \\
& & & & & & & \ddots & \ddots & \\
& & & & & & & & & I
\end{pmatrix}
\tag{II.68}
$$

and

$$
\mathbf{b}^c = \begin{pmatrix}
b^{(0)_c}(t_1) + b^{(1)_c}(t_1) + \cdots + b^{(p)_c}(t_1) \\
Y^{(1)}(t_1) \\
\vdots \\
Y^{(p-1)}(t_1) \\
b^{(0)_c}(t_2) + b^{(1)_c}(t_2) + \cdots + b^{(p)_c}(t_2) \\
Y^{(1)}(t_2) \\
\vdots \\
\vdots \\
b^{(0)_c}(t_M) + b^{(1)_c}(t_M) + \cdots + b^{(p-1)_c}(t_M) \\
Y^{(1)}(t_M) \\
\vdots \\
Y^{(p-1)}(t_M) \\
X^c(0)
\end{pmatrix} .
\tag{II.69}
$$

The initial conditions are given by

$$\boldsymbol{X^c}(0) = \begin{pmatrix} x^c(0) \\ (x^c)^2(0) \\ \vdots \\ (x^c)^N(0) \end{pmatrix}.$$

(II.70)

One important feature of the vector defined in Eq. (II.69) is that it is a sparse vector. Specifically, as shown in Eq. (II.59), each vector $b^{(m)_c}$ contains *only one nonzero entry* within an $N$-dimensional space. Here, $N$ represents the truncation parameter, which is assumed to be significantly larger than $p$, the number of interpolation points chosen manually as $N \gg p$. To conclude, we would like to study the query complexity of this algorithm. As a first step, we observe that the sparsity of matrix $\mathbf{M^c}$ is given by

$$s_{\mathbf{M^c}} = (p+1) \cdot J + 2.$$

(II.71)

Now we are once more in the position to invoke the state-of-the-art QLSS [21–23] or the state-of-the-art LCHS algorithm [18, 29] to solve this linear system.

### C. Uploading and downloading problems

In this section, we address the uploading and downloading problems in quantum computing, particularly in the context of *quantum linear system solvers* (QLSS) and their applications to *diffusion probabilistic models* (DPMs). A key feature of our framework is that the inputs and outputs on the quantum device are sparse, which means we do not necessarily require access to *quantum random access memory* (QRAM). This eliminates one of the major challenges in implementing quantum algorithms at scale, as QRAM relies on non-local quantum controls that are experimentally difficult to achieve [32, 33]. Additionally, the non-local nature of QRAM imposes physical size limitations governed by the principles of special relativity [34]. By focusing on sparse data representations, our algorithm sidesteps these challenges, ensuring both scalability and practicality.

The sparsity of our setup aligns naturally with sparse training methods, which are widely studied in classical machine learning for reducing computational overhead while maintaining model performance [35–38]. Sparse training setups allow quantum algorithms to process classical data efficiently without relying on QRAM [13, 39]. In our approach, similar to that presented in Ref. [13], the interface between classical and quantum processors only needs to be established once for $M$ iterations, significantly reducing the burden of data transmission between the classical and quantum systems. This characteristic makes the sparse training setup particularly advantageous for near-term quantum devices, where minimizing classical-to-quantum communication overhead is critical.

While our focus is on sparse training, dense training scenarios are also relevant in certain applications. For example, tasks like high-resolution image generation often require large datasets and densely parameterized models to learn intricate and detailed data distributions [1–3]. In such cases, QRAM may be necessary to efficiently encode non-sparse classical data into a quantum computer. QRAM has been proposed as a potential solution for data uploading [32, 40–42], and some devices have demonstrated polynomially controlled error scaling [40]. However, implementing QRAM in practice remains highly challenging, particularly in large-scale, fault-tolerant regimes.

The downloading problem can be more challenging than the uploading problem. After running the quantum algorithm, we obtain the quantum states defined in Eq. (III.3) and Eq. (III.7). Downloading these resulting quantum states involves performing state tomography [43–45], which can demand substantial resources for generic dense quantum states. However, the process becomes much more efficient for sparse states, as the sparsity significantly reduces the number of measurements required to reconstruct the state.

#### 1. Tomographic recovery and sparsity

In this section, we provide a brief overview of tomographic recovery [46] and its significance in sparse training following Ref. [13]. The implementation of our quantum algorithms relies on two key aspects of sparsity: the sparsity of the vectors $x_\lambda$ and the sparsity of the QCM matrix $A$. The sparsity of the QCM matrix arises naturally from the DPM architecture and Taylor expansion, while the sparsity of weight matrices is a direct consequence of the sparse training approach. Here, we focus primarily on the sparsity of the output vector, denoting by $r$ the number of non-zero entries in a vector within a Hilbert space of dimension $N$. The state vectors to be recovered tomographically in this step are of the form

$$|x\rangle = \sum_{\alpha=1}^{r} v_\alpha |p_\alpha\rangle,$$

(II.72)

normalized as

$$\sum_{\alpha=1}^{r} |v_\alpha|^2 = 1,$$

(II.73)

where each $|p_\alpha\rangle$ is a computational basis vector of a system of $m = \log_2(n)$ qubits. In other words, the state is $r$-sparse, meaning it contains at most $r$ non-zero coefficients. The first step involves identifying which computational basis vectors are present and, consequently, which coefficients are non-zero, as the sparsity pattern is not initially known. This task can be framed as an instance of the *coupon collector's problem* (see, for example, Ref. [47]). The algorithm for this step proceeds as follows: projective measurements are performed in the computational Pauli-$Z$ basis, and the resulting patterns are recorded. This process is repeated until all $r$ distinct patterns are observed. The classical *coupon collector's problem* corresponds precisely to the scenario of a state vector $|x\rangle = r^{-1/2} \sum_{\alpha=1}^{r} |p_\alpha\rangle$, where each pattern appears with equal probability due to the uniform superposition. In this case, the expected number of measurements required scales as $\mathcal{O}(r \log r)$. For the state vector specified in Eq. (II.72), where the weights associated with each pattern differ, the situation generalizes to a *weighted coupon collector's problem*. The expected number of repetitions needed to observe all patterns in this context is also well established [13, 47]. The expected number of steps is given by

$$\int_0^{+\infty} \left(1 - \prod_{\alpha=1}^{r} \left(1 - e^{-|v_\alpha|^2 y}\right)\right) dy,$$

(II.74)

which is equal to

$$\sum_{q=0}^{r-1} (-1)^{r-1-q} \sum_{|J|=q} \frac{1}{1-P_j},$$

(II.75)

where $r$ denotes the number of coupons to be collected and $P_J$ the probability of getting any coupon in the set of coupons $J$. It is worth noting that the minimum expected number of measurements is achieved when all patterns are equally probable. However, a subtle issue arises in the presence of very small non-zero weights. As these weights approach zero, the expected number of measurements required to observe all patterns – including those corresponding to the smallest weights – diverge. To address this, one can focus solely on the significant weights, defined as those exceeding a certain threshold $\delta > 0$. By considering only these significant weights, an efficient algorithm can be devised to approximate the result while disregarding negligible contributions. As detailed analysis demonstrates, this approach still requires $\mathcal{O}(r \log r)$ measurements for significant weights, ensuring computational efficiency.

After identifying the at most $r$ relevant computational basis vectors corresponding to an $r$-sparse weight state, we can proceed with a variant of state tomography to reconstruct the complete pure quantum state, including its phases. To achieve this, we propose leveraging a modified version of the *shadow estimation scheme* introduced in Ref. [44], which utilizes random quantum circuits built from the $n$-qubit Clifford group $\mathcal{C}_n$. For the physical implementation of this approach, approximately $\mathcal{O}(m^2)$ entangling gates are required. Once the random Clifford circuits has been applied, computational basis measurements are performed, yielding outcomes $o$ that correspond to computational basis vectors $|o\rangle$ of the $m$-qubit system. For the input state vector $|x\rangle\langle x|$, the random quantum channel applied through these Clifford circuits produces the output

$$\mathcal{M}(|x\rangle\langle x|) = \mathbb{E}\left(U^\dagger |o\rangle\langle o|U\right),$$

(II.76)

where $U \in \mathcal{C}_m$ is an element from the $m$-qubit Clifford group. To recover the original state $|x\rangle\langle x|$, we can classically compute the inverse of this process as

$$|x\rangle\langle x| = \mathbb{E}\left(\mathcal{M}^{-1}(U^\dagger |o\rangle\langle o|U)\right),$$

(II.77)

as long as sufficiently many samples are available. This requires computing the overlaps of computational basis vectors with stabilizer states, which can be efficiently done classically with a polynomial computational effort in $m$, and the inverse map is

$$\mathcal{M}^{-1}(X) = (2^m + 1)X - I.$$

(II.78)

For a state vector of the form given in Eq. (II.72), one can take samples of random Clifford circuits applied to $|x\rangle$ and sample from the output distribution resulting from computational basis measurements, thus creating a classical shadow. Using these data, we can estimate the elements $v_\alpha$ for $\alpha = 1, \ldots, r$. The overall effort required for this procedure is $\mathcal{O}(m^2 r^3/\epsilon^2)$ in the random measurement procedure, with a recovery error bounded by the 1-norm error $\epsilon > 0$, ignoring logarithmic factors. Additionally, this process involves a polynomial classical effort in $m$, which arises from computing overlaps of stabilizer states with computational basis states. The depth of the Clifford circuits can be further reduced [48].

In the actual DPM process similar to the general machine learning process described in Ref. [13], the ratio of pruned parameters in sparse training $\delta_{\mathrm{tr}}(N)$ scales with the size of the model $N$. In this context, the weight sparsity is given by $r = N(1 - \delta_{\mathrm{tr}}(N))$. For this approach to be feasible, we expect that $r$ remains constant as $N$ increases. This implies that

$$\delta_{\mathrm{tr}}(N) = 1 - \mathcal{O}\left(\frac{1}{N}\right). \tag{II.79}$$

This condition ensures that the cost of our algorithm exhibits poly-logarithmic dependence on $N$. Finally, in the case of dense input vectors, the tomographic algorithm will focus on reading the $r$ vector components with the largest norms, which can be seen as a natural pruning method for subsequent classical processes. Here, we assume a generic dense training approach, where the input dense weight vectors can be uploaded via QRAM. The downloading process can remain efficient as long as $r$ does not scale with $N$, and this can be interpreted as a hybrid quantum-classical pruning method.

## III. THEOREMS

In this section, building on our previous discussions on quantum ODE solvers and quantum tomography, we present two theorems that characterize the DPM-solver and UniPC approaches, respectively. Drawing from the theorems and quantum algorithms presented in Refs. [13, 21–23], we establish a gate complexity bound for the quantum Carleman algorithm described in this section. It is important to note that the theorems in Refs. [21–23] do not explicitly account for the dependence on the sparsity $s_M$ of the matrix $M$ and the system size $N$, as these dependencies are implicitly encapsulated in the cost functions $\mathbf{Cost}(\mathcal{O}_M)$. A detailed investigation of this dependence can be found in Lemma 48 of Ref. [24].

**Theorem III.1** (Gate complexity bound for the quantum Carleman algorithm (DPM-solver)). *Given a DPM-solver with denoising function $\epsilon_\theta(x_\lambda, \lambda)$ and its higher-order derivatives expanded to $J$-th order Taylor approximation in terms of $x_\lambda$, it is possible to formulate a system of linear equations: $M\boldsymbol{Y} = \boldsymbol{b}$, after truncating the system at $N$, as shown in Eq. (II.26), where $A$ is the quantum Carleman matrix for DPM-solver, defined in Eq. (II.23). $M$ is a sparse matrix with sparsity*

$$s_M \leq J + 1. \tag{III.1}$$

*$M/\alpha_M$ is block encoded by $\mathcal{O}_M$ with normalization factor $\alpha_M \geq \|M\|$. Let $|\boldsymbol{b}\rangle$ be the initial state vector prepared by the oracle $\mathcal{O}_{\boldsymbol{b}}$. Suppose that the solution norm $\|M^{-1}|\boldsymbol{b}\rangle\|$, and hence the success amplitude*

$$\sqrt{p_{succ}} = \frac{\|M^{-1}|\boldsymbol{b}\rangle\|}{\alpha_{M^{-1}}}, \tag{III.2}$$

*can be estimated to a constant multiplicative accuracy. Then the quantum state vector*

$$\frac{M^{-1}|\boldsymbol{b}\rangle}{\|M^{-1}|\boldsymbol{b}\rangle\|}, \tag{III.3}$$

*can be prepared to accuracy $\epsilon > 0$ and success probability $> 1/2$ with query complexity*

$$\mathcal{O}\left(J\kappa \cdot \mathrm{polylog}\left(\frac{N}{\epsilon}\right)\right), \tag{III.4}$$

*where $\alpha_{M^{-1}} \geq \|M^{-1}\|$ is a norm upper bound on the inverse matrix, and $\kappa = \alpha_M \alpha_{M^{-1}}$ is an upper bound on the spectral condition number. Moreover, assuming that the output weight vectors are $r$-sparse and $m := \log_2(N)$, the tomographic cost to transfer quantum states to classical devices is $\mathcal{O}(m^2 r^3/\epsilon^2)$, ignoring logarithmic factors. The algorithm does not account for the state preparation cost required to initialize the weight vector on the quantum device, which remains computationally efficient within the sparse training framework.*

In addition to this, by basically the same logic, we also have a gate complexity theorem for the quantum algorithm for UniPC.

**Theorem III.2** (Gate complexity bound for the quantum Carleman algorithm (UniPC-$p$)). *Given a unified predictor-corrector (UniPC-p) framework for fast sampling of DPMs with denoising function $\epsilon_\theta(x_\lambda, \lambda)$ expanded to $J$-th order Taylor approximation in terms of $x_\lambda$, it is possible to formulate a system of linear equations: $\mathsf{M}^c\mathsf{Z} = \mathbf{b}^c$, after truncating the system at $N$, where $\mathsf{M}^c$ and $\mathbf{b}^c$ are defined respectively in Eq. (II.68) and Eq. (II.69). The Carleman matrices for UniPC at all iteration and interpolation time steps are given in Eq. (II.58). $\mathsf{M}^c$ is a sparse matrix with sparsity*

$$s_{\mathsf{M}^c} \leq (p+1) \cdot J + 2. \tag{III.5}$$


}



\begin{thebibliography}{49}%
\makeatletter
\providecommand \@ifxundefined [1]{%
 \@ifx{#1\undefined}
}%
\providecommand \@ifnum [1]{%
 \ifnum #1\expandafter \@firstoftwo
 \else \expandafter \@secondoftwo
 \fi
}%
\providecommand \@ifx [1]{%
 \ifx #1\expandafter \@firstoftwo
 \else \expandafter \@secondoftwo
 \fi
}%
\providecommand \natexlab [1]{#1}%
\providecommand \enquote  [1]{``#1''}%
\providecommand \bibnamefont  [1]{#1}%
\providecommand \bibfnamefont [1]{#1}%
\providecommand \citenamefont [1]{#1}%
\providecommand \href@noop [0]{\@secondoftwo}%
\providecommand \href [0]{\begingroup \@sanitize@url \@href}%
\providecommand \@href[1]{\@@startlink{#1}\@@href}%
\providecommand \@@href[1]{\endgroup#1\@@endlink}%
\providecommand \@sanitize@url [0]{\catcode `\\12\catcode `\$12\catcode
  `\&12\catcode `\#12\catcode `\^12\catcode `\_12\catcode `\%12\relax}%
\providecommand \@@startlink[1]{}%
\providecommand \@@endlink[0]{}%
\providecommand \url  [0]{\begingroup\@sanitize@url \@url }%
\providecommand \@url [1]{\endgroup\@href {#1}{\urlprefix }}%
\providecommand \urlprefix  [0]{URL }%
\providecommand \Eprint [0]{\href }%
\providecommand \doibase [0]{https://doi.org/}%
\providecommand \selectlanguage [0]{\@gobble}%
\providecommand \bibinfo  [0]{\@secondoftwo}%
\providecommand \bibfield  [0]{\@secondoftwo}%
\providecommand \translation [1]{[#1]}%
\providecommand \BibitemOpen [0]{}%
\providecommand \bibitemStop [0]{}%
\providecommand \bibitemNoStop [0]{.\EOS\space}%
\providecommand \EOS [0]{\spacefactor3000\relax}%
\providecommand \BibitemShut  [1]{\csname bibitem#1\endcsname}%
\let\auto@bib@innerbib\@empty
\bibitem [{\citenamefont {Chang}\ \emph {et~al.}(2023)\citenamefont {Chang},
  \citenamefont {Koulieris},\ and\ \citenamefont {Shum}}]{DiffusionModels1}%
  \BibitemOpen
  \bibfield  {author} {\bibinfo {author} {\bibfnamefont {Z.}~\bibnamefont
  {Chang}}, \bibinfo {author} {\bibfnamefont {G.~A.}\ \bibnamefont
  {Koulieris}},\ and\ \bibinfo {author} {\bibfnamefont {H.~P.~H.}\ \bibnamefont
  {Shum}},\ }\href {https://arxiv.org/abs/2306.04542} {\bibinfo {title} {On the
  design fundamentals of diffusion models: A survey}} (\bibinfo {year}
  {2023}),\ \Eprint {https://arxiv.org/abs/2306.04542} {arXiv:2306.04542}
  \BibitemShut {NoStop}%
\bibitem [{\citenamefont {Croitoru}\ \emph {et~al.}(2023)\citenamefont
  {Croitoru}, \citenamefont {Hondru}, \citenamefont {Ionescu},\ and\
  \citenamefont {Shah}}]{DiffusionModels2}%
  \BibitemOpen
  \bibfield  {author} {\bibinfo {author} {\bibfnamefont {F.-A.}\ \bibnamefont
  {Croitoru}}, \bibinfo {author} {\bibfnamefont {V.}~\bibnamefont {Hondru}},
  \bibinfo {author} {\bibfnamefont {R.~T.}\ \bibnamefont {Ionescu}},\ and\
  \bibinfo {author} {\bibfnamefont {M.}~\bibnamefont {Shah}},\ }\bibfield
  {title} {\bibinfo {title} {Diffusion models in vision: A survey},\ }\href
  {https://doi.org/10.1109/tpami.2023.3261988} {\bibfield  {journal} {\bibinfo
  {journal} {IEEE Trans. Patt. Ana. Mach. Intel.}\ }\textbf {\bibinfo {volume}
  {45}},\ \bibinfo {pages} {10850–10869} (\bibinfo {year}
  {2023})}\BibitemShut {NoStop}%
\bibitem [{\citenamefont {Song}\ \emph {et~al.}(2021)\citenamefont {Song},
  \citenamefont {Sohl-Dickstein}, \citenamefont {Kingma}, \citenamefont
  {Kumar}, \citenamefont {Ermon},\ and\ \citenamefont
  {Poole}}]{DiffusionModels3}%
  \BibitemOpen
  \bibfield  {author} {\bibinfo {author} {\bibfnamefont {Y.}~\bibnamefont
  {Song}}, \bibinfo {author} {\bibfnamefont {J.}~\bibnamefont
  {Sohl-Dickstein}}, \bibinfo {author} {\bibfnamefont {D.~P.}\ \bibnamefont
  {Kingma}}, \bibinfo {author} {\bibfnamefont {A.}~\bibnamefont {Kumar}},
  \bibinfo {author} {\bibfnamefont {S.}~\bibnamefont {Ermon}},\ and\ \bibinfo
  {author} {\bibfnamefont {B.}~\bibnamefont {Poole}},\ }\href
  {https://arxiv.org/abs/2011.13456} {\bibinfo {title} {Score-based generative
  modeling through stochastic differential equations}} (\bibinfo {year}
  {2021}),\ \Eprint {https://arxiv.org/abs/2011.13456} {arXiv:2011.13456}
  \BibitemShut {NoStop}%
\bibitem [{\citenamefont {Ho}\ \emph {et~al.}(2020)\citenamefont {Ho},
  \citenamefont {Jain},\ and\ \citenamefont {Abbeel}}]{ho2020denoising}%
  \BibitemOpen
  \bibfield  {author} {\bibinfo {author} {\bibfnamefont {J.}~\bibnamefont
  {Ho}}, \bibinfo {author} {\bibfnamefont {A.}~\bibnamefont {Jain}},\ and\
  \bibinfo {author} {\bibfnamefont {P.}~\bibnamefont {Abbeel}},\ }\href
  {https://arxiv.org/abs/2006.11239} {\bibinfo {title} {Denoising diffusion
  probabilistic models}} (\bibinfo {year} {2020}),\ \Eprint
  {https://arxiv.org/abs/2006.11239} {arXiv:2006.11239} \BibitemShut {NoStop}%
\bibitem [{\citenamefont {Creswell}\ \emph {et~al.}(2017)\citenamefont
  {Creswell}, \citenamefont {White}, \citenamefont {Dumoulin},\ and\
  \citenamefont {Arulkumaran}}]{GANS}%
  \BibitemOpen
  \bibfield  {author} {\bibinfo {author} {\bibfnamefont {A.}~\bibnamefont
  {Creswell}}, \bibinfo {author} {\bibfnamefont {T.}~\bibnamefont {White}},
  \bibinfo {author} {\bibfnamefont {V.}~\bibnamefont {Dumoulin}},\ and\
  \bibinfo {author} {\bibfnamefont {K.}~\bibnamefont {Arulkumaran}},\
  }\bibfield  {title} {\bibinfo {title} {Generative adversarial networks: An
  overview},\ }\href {https://doi.org/10.1109/MSP.2017.2765202} {\bibfield
  {journal} {\bibinfo  {journal} {IEEE Sig. Proc. Mag.}\ }\textbf {\bibinfo
  {volume} {35}},\ \bibinfo {pages} {53} (\bibinfo {year} {2017})}\BibitemShut
  {NoStop}%
\bibitem [{\citenamefont {Kingma}\ and\ \citenamefont {Welling}(2019)}]{VAE}%
  \BibitemOpen
  \bibfield  {author} {\bibinfo {author} {\bibfnamefont {D.~P.}\ \bibnamefont
  {Kingma}}\ and\ \bibinfo {author} {\bibfnamefont {M.}~\bibnamefont
  {Welling}},\ }\bibfield  {title} {\bibinfo {title} {An introduction to
  variational autoencoders},\ }\href {https://doi.org/10.1561/2200000056}
  {\bibfield  {journal} {\bibinfo  {journal} {Found. Trends Mach. Learn.}\
  }\textbf {\bibinfo {volume} {12}},\ \bibinfo {pages} {307–392} (\bibinfo
  {year} {2019})}\BibitemShut {NoStop}%
\bibitem [{\citenamefont {Preskill}(1998)}]{preskill1998lecture}%
  \BibitemOpen
  \bibfield  {author} {\bibinfo {author} {\bibfnamefont {J.}~\bibnamefont
  {Preskill}},\ }\bibfield  {title} {\bibinfo {title} {Lecture notes for
  physics 229: Quantum information and computation},\ }\href
  {https://www.preskill.caltech.edu/ph229} {\bibfield  {journal} {\bibinfo
  {journal} {California Institute of Technology}\ }\textbf {\bibinfo {volume}
  {16}},\ \bibinfo {pages} {1} (\bibinfo {year} {1998})}\BibitemShut {NoStop}%
\bibitem [{\citenamefont {Shor}(1999)}]{shor1999polynomial}%
  \BibitemOpen
  \bibfield  {author} {\bibinfo {author} {\bibfnamefont {P.~W.}\ \bibnamefont
  {Shor}},\ }\bibfield  {title} {\bibinfo {title} {Polynomial-time algorithms
  for prime factorization and discrete logarithms on a quantum computer},\
  }\href {https://doi.org/10.1137/S0036144598347011} {\bibfield  {journal}
  {\bibinfo  {journal} {SIAM Rev.}\ }\textbf {\bibinfo {volume} {41}},\
  \bibinfo {pages} {303} (\bibinfo {year} {1999})}\BibitemShut {NoStop}%
\bibitem [{\citenamefont {Grover}(1996)}]{grover1996fast}%
  \BibitemOpen
  \bibfield  {author} {\bibinfo {author} {\bibfnamefont {L.~K.}\ \bibnamefont
  {Grover}},\ }\bibfield  {title} {\bibinfo {title} {A fast quantum mechanical
  algorithm for database search},\ }in\ \href@noop {} {\emph {\bibinfo
  {booktitle} {Proceedings of the twenty-eighth annual ACM symposium on Theory
  of computing}}}\ (\bibinfo {year} {1996})\ pp.\ \bibinfo {pages} {212--219},\
  \bibinfo {note}
  {\href{https://arxiv.org/abs/quant-ph/9605043}{arXiv:quant-ph/9605043}}\BibitemShut
  {NoStop}%
\bibitem [{\citenamefont {Lloyd}(1996)}]{lloyd1996universal}%
  \BibitemOpen
  \bibfield  {author} {\bibinfo {author} {\bibfnamefont {S.}~\bibnamefont
  {Lloyd}},\ }\bibfield  {title} {\bibinfo {title} {Universal quantum
  simulators},\ }\href {https://doi.org/10.1126/science.273.5278.1073}
  {\bibfield  {journal} {\bibinfo  {journal} {Science}\ }\textbf {\bibinfo
  {volume} {273}},\ \bibinfo {pages} {1073} (\bibinfo {year}
  {1996})}\BibitemShut {NoStop}%
\bibitem [{\citenamefont {Hangleiter}\ and\ \citenamefont
  {Eisert}(2023)}]{SupremacyReview}%
  \BibitemOpen
  \bibfield  {author} {\bibinfo {author} {\bibfnamefont {D.}~\bibnamefont
  {Hangleiter}}\ and\ \bibinfo {author} {\bibfnamefont {J.}~\bibnamefont
  {Eisert}},\ }\bibfield  {title} {\bibinfo {title} {Computational advantage of
  quantum random sampling},\ }\href
  {https://doi.org/10.1103/RevModPhys.95.035001} {\bibfield  {journal}
  {\bibinfo  {journal} {Rev. Mod. Phys.}\ }\textbf {\bibinfo {volume} {95}},\
  \bibinfo {pages} {035001} (\bibinfo {year} {2023})}\BibitemShut {NoStop}%
\bibitem [{\citenamefont {Wang}\ and\ \citenamefont {Liu}(2024)}]{Wang_2024}%
  \BibitemOpen
  \bibfield  {author} {\bibinfo {author} {\bibfnamefont {Y.}~\bibnamefont
  {Wang}}\ and\ \bibinfo {author} {\bibfnamefont {J.}~\bibnamefont {Liu}},\
  }\bibfield  {title} {\bibinfo {title} {{A comprehensive review of quantum
  machine learning: from NISQ to fault tolerance}},\ }\href
  {https://doi.org/10.1088/1361-6633/ad7f69} {\bibfield  {journal} {\bibinfo
  {journal} {Rep. Prog. Phys.}\ }\textbf {\bibinfo {volume} {87}},\ \bibinfo
  {pages} {116402} (\bibinfo {year} {2024})}\BibitemShut {NoStop}%
\bibitem [{\citenamefont {Harrow}\ \emph {et~al.}(2009)\citenamefont {Harrow},
  \citenamefont {Hassidim},\ and\ \citenamefont {Lloyd}}]{Harrow_2009}%
  \BibitemOpen
  \bibfield  {author} {\bibinfo {author} {\bibfnamefont {A.~W.}\ \bibnamefont
  {Harrow}}, \bibinfo {author} {\bibfnamefont {A.}~\bibnamefont {Hassidim}},\
  and\ \bibinfo {author} {\bibfnamefont {S.}~\bibnamefont {Lloyd}},\ }\bibfield
   {title} {\bibinfo {title} {Quantum algorithm for linear systems of
  equations},\ }\href {https://doi.org/10.1103/physrevlett.103.150502}
  {\bibfield  {journal} {\bibinfo  {journal} {Phys. Rev. Lett.}\ }\textbf
  {\bibinfo {volume} {103}},\ \bibinfo {pages} {150502} (\bibinfo {year}
  {2009})}\BibitemShut {NoStop}%
\bibitem [{\citenamefont {Preskill}(2025)}]{Megaquop}%
  \BibitemOpen
  \bibfield  {author} {\bibinfo {author} {\bibfnamefont {J.}~\bibnamefont
  {Preskill}},\ }\href {https://arxiv.org/abs/2502.17368} {\bibinfo {title}
  {{Beyond NISQ: The megaquop machine}}} (\bibinfo {year} {2025}),\ \Eprint
  {https://arxiv.org/abs/2502.17368} {arXiv:2502.17368} \BibitemShut {NoStop}%
\bibitem [{\citenamefont {Eisert}\ and\ \citenamefont
  {Preskill}(2025)}]{MindTheGaps}%
  \BibitemOpen
  \bibfield  {author} {\bibinfo {author} {\bibfnamefont {J.}~\bibnamefont
  {Eisert}}\ and\ \bibinfo {author} {\bibfnamefont {J.}~\bibnamefont
  {Preskill}},\ }\href {https://arxiv.org/abs/2510.19928} {\bibinfo {title}
  {Mind the gaps: The fraught road to quantum advantage}} (\bibinfo {year}
  {2025}),\ \Eprint {https://arxiv.org/abs/2510.19928} {arXiv:2510.19928}
  \BibitemShut {NoStop}%
\bibitem [{\citenamefont {Biamonte}\ \emph {et~al.}(2017)\citenamefont
  {Biamonte}, \citenamefont {Wittek}, \citenamefont {Pancotti}, \citenamefont
  {Rebentrost}, \citenamefont {Wiebe},\ and\ \citenamefont
  {Lloyd}}]{biamonte2017quantum}%
  \BibitemOpen
  \bibfield  {author} {\bibinfo {author} {\bibfnamefont {J.}~\bibnamefont
  {Biamonte}}, \bibinfo {author} {\bibfnamefont {P.}~\bibnamefont {Wittek}},
  \bibinfo {author} {\bibfnamefont {N.}~\bibnamefont {Pancotti}}, \bibinfo
  {author} {\bibfnamefont {P.}~\bibnamefont {Rebentrost}}, \bibinfo {author}
  {\bibfnamefont {N.}~\bibnamefont {Wiebe}},\ and\ \bibinfo {author}
  {\bibfnamefont {S.}~\bibnamefont {Lloyd}},\ }\bibfield  {title} {\bibinfo
  {title} {Quantum machine learning},\ }\href
  {https://doi.org/10.1038/nature23474} {\bibfield  {journal} {\bibinfo
  {journal} {Nature}\ }\textbf {\bibinfo {volume} {549}},\ \bibinfo {pages}
  {195} (\bibinfo {year} {2017})}\BibitemShut {NoStop}%
\bibitem [{\citenamefont {Carleo}\ \emph {et~al.}(2019)\citenamefont {Carleo},
  \citenamefont {Cirac}, \citenamefont {Cranmer}, \citenamefont {Daudet},
  \citenamefont {Schuld}, \citenamefont {Tishby}, \citenamefont
  {Vogt-Maranto},\ and\ \citenamefont {Zdeborov\'a}}]{RevModPhys.91.045002}%
  \BibitemOpen
  \bibfield  {author} {\bibinfo {author} {\bibfnamefont {G.}~\bibnamefont
  {Carleo}}, \bibinfo {author} {\bibfnamefont {I.}~\bibnamefont {Cirac}},
  \bibinfo {author} {\bibfnamefont {K.}~\bibnamefont {Cranmer}}, \bibinfo
  {author} {\bibfnamefont {L.}~\bibnamefont {Daudet}}, \bibinfo {author}
  {\bibfnamefont {M.}~\bibnamefont {Schuld}}, \bibinfo {author} {\bibfnamefont
  {N.}~\bibnamefont {Tishby}}, \bibinfo {author} {\bibfnamefont
  {L.}~\bibnamefont {Vogt-Maranto}},\ and\ \bibinfo {author} {\bibfnamefont
  {L.}~\bibnamefont {Zdeborov\'a}},\ }\bibfield  {title} {\bibinfo {title}
  {Machine learning and the physical sciences},\ }\href
  {https://doi.org/10.1103/RevModPhys.91.045002} {\bibfield  {journal}
  {\bibinfo  {journal} {Rev. Mod. Phys.}\ }\textbf {\bibinfo {volume} {91}},\
  \bibinfo {pages} {045002} (\bibinfo {year} {2019})}\BibitemShut {NoStop}%
\bibitem [{\citenamefont {McClean}\ \emph {et~al.}(2016)\citenamefont
  {McClean}, \citenamefont {Romero}, \citenamefont {Babbush},\ and\
  \citenamefont {Aspuru-Guzik}}]{McClean_2016}%
  \BibitemOpen
  \bibfield  {author} {\bibinfo {author} {\bibfnamefont {J.~R.}\ \bibnamefont
  {McClean}}, \bibinfo {author} {\bibfnamefont {J.}~\bibnamefont {Romero}},
  \bibinfo {author} {\bibfnamefont {R.}~\bibnamefont {Babbush}},\ and\ \bibinfo
  {author} {\bibfnamefont {A.}~\bibnamefont {Aspuru-Guzik}},\ }\bibfield
  {title} {\bibinfo {title} {The theory of variational hybrid quantum-classical
  algorithms},\ }\href {https://doi.org/10.1088/1367-2630/18/2/023023}
  {\bibfield  {journal} {\bibinfo  {journal} {New J. Phys.}\ }\textbf {\bibinfo
  {volume} {18}},\ \bibinfo {pages} {023023} (\bibinfo {year}
  {2016})}\BibitemShut {NoStop}%
\bibitem [{\citenamefont {Sweke}\ \emph {et~al.}(2021)\citenamefont {Sweke},
  \citenamefont {Seifert}, \citenamefont {Hangleiter},\ and\ \citenamefont
  {Eisert}}]{PACLearning}%
  \BibitemOpen
  \bibfield  {author} {\bibinfo {author} {\bibfnamefont {R.}~\bibnamefont
  {Sweke}}, \bibinfo {author} {\bibfnamefont {J.-P.}\ \bibnamefont {Seifert}},
  \bibinfo {author} {\bibfnamefont {D.}~\bibnamefont {Hangleiter}},\ and\
  \bibinfo {author} {\bibfnamefont {J.}~\bibnamefont {Eisert}},\ }\bibfield
  {title} {\bibinfo {title} {On the quantum versus classical learnability of
  discrete distributions},\ }\href {https://doi.org/10.22331/q-2021-03-23-417}
  {\bibfield  {journal} {\bibinfo  {journal} {Quantum}\ }\textbf {\bibinfo
  {volume} {5}},\ \bibinfo {pages} {417} (\bibinfo {year} {2021})}\BibitemShut
  {NoStop}%
\bibitem [{\citenamefont {Pirnay}\ \emph {et~al.}(2023)\citenamefont {Pirnay},
  \citenamefont {Sweke}, \citenamefont {Eisert},\ and\ \citenamefont
  {Seifert}}]{DensityModelling}%
  \BibitemOpen
  \bibfield  {author} {\bibinfo {author} {\bibfnamefont {N.}~\bibnamefont
  {Pirnay}}, \bibinfo {author} {\bibfnamefont {R.}~\bibnamefont {Sweke}},
  \bibinfo {author} {\bibfnamefont {J.}~\bibnamefont {Eisert}},\ and\ \bibinfo
  {author} {\bibfnamefont {J.-P.}\ \bibnamefont {Seifert}},\ }\bibfield
  {title} {\bibinfo {title} {A super-polynomial quantum-classical separation
  for density modelling},\ }\href {https://doi.org/10.1103/PhysRevA.107.042416}
  {\bibfield  {journal} {\bibinfo  {journal} {Phys. Rev. A}\ }\textbf {\bibinfo
  {volume} {107}},\ \bibinfo {pages} {042416} (\bibinfo {year}
  {2023})}\BibitemShut {NoStop}%
\bibitem [{\citenamefont {Liu}\ \emph {et~al.}(2021{\natexlab{a}})\citenamefont
  {Liu}, \citenamefont {Arunachalam},\ and\ \citenamefont {Temme}}]{TemmeML}%
  \BibitemOpen
  \bibfield  {author} {\bibinfo {author} {\bibfnamefont {Y.}~\bibnamefont
  {Liu}}, \bibinfo {author} {\bibfnamefont {S.}~\bibnamefont {Arunachalam}},\
  and\ \bibinfo {author} {\bibfnamefont {K.}~\bibnamefont {Temme}},\ }\bibfield
   {title} {\bibinfo {title} {A rigorous and robust quantum speed-up in
  supervised machine learning},\ }\href
  {https://doi.org/10.1038/s41567-021-01287-z} {\bibfield  {journal} {\bibinfo
  {journal} {Nature Phys.}\ }\textbf {\bibinfo {volume} {17}},\ \bibinfo
  {pages} {1013} (\bibinfo {year} {2021}{\natexlab{a}})}\BibitemShut {NoStop}%
\bibitem [{\citenamefont {Pirnay}\ \emph {et~al.}(2024)\citenamefont {Pirnay},
  \citenamefont {Jerbi}, \citenamefont {Seifert},\ and\ \citenamefont
  {Eisert}}]{ShortCircuitsLearning}%
  \BibitemOpen
  \bibfield  {author} {\bibinfo {author} {\bibfnamefont {N.}~\bibnamefont
  {Pirnay}}, \bibinfo {author} {\bibfnamefont {S.}~\bibnamefont {Jerbi}},
  \bibinfo {author} {\bibfnamefont {J.~P.}\ \bibnamefont {Seifert}},\ and\
  \bibinfo {author} {\bibfnamefont {J.}~\bibnamefont {Eisert}},\ }\href
  {https://arxiv.org/abs/2411.15548} {\bibinfo {title} {An unconditional
  distribution learning advantage with shallow quantum circuits}} (\bibinfo
  {year} {2024}),\ \Eprint {https://arxiv.org/abs/2411.15548}
  {arXiv:2411.15548} \BibitemShut {NoStop}%
\bibitem [{\citenamefont {Schuld}\ and\ \citenamefont
  {Killoran}(2022)}]{PRXQuantum.3.030101}%
  \BibitemOpen
  \bibfield  {author} {\bibinfo {author} {\bibfnamefont {M.}~\bibnamefont
  {Schuld}}\ and\ \bibinfo {author} {\bibfnamefont {N.}~\bibnamefont
  {Killoran}},\ }\bibfield  {title} {\bibinfo {title} {Is quantum advantage the
  right goal for quantum machine learning?},\ }\href
  {https://doi.org/10.1103/PRXQuantum.3.030101} {\bibfield  {journal} {\bibinfo
   {journal} {PRX Quantum}\ }\textbf {\bibinfo {volume} {3}},\ \bibinfo {pages}
  {030101} (\bibinfo {year} {2022})}\BibitemShut {NoStop}%
\bibitem [{\citenamefont {Zimborás}\ \emph {et~al.}(2025)\citenamefont
  {Zimborás}, \citenamefont {Koczor}, \citenamefont {Holmes}, \citenamefont
  {Borrelli}, \citenamefont {Gilyén}, \citenamefont {Huang}, \citenamefont
  {Cai}, \citenamefont {Acín}, \citenamefont {Aolita}, \citenamefont {Banchi},
  \citenamefont {Brandão}, \citenamefont {Cavalcanti}, \citenamefont {Cubitt},
  \citenamefont {Filippov}, \citenamefont {García-Pérez}, \citenamefont
  {Goold}, \citenamefont {Kálmán}, \citenamefont {Kyoseva}, \citenamefont
  {Rossi}, \citenamefont {Sokolov}, \citenamefont {Tavernelli},\ and\
  \citenamefont {Maniscalco}}]{Myths}%
  \BibitemOpen
  \bibfield  {author} {\bibinfo {author} {\bibfnamefont {Z.}~\bibnamefont
  {Zimborás}}, \bibinfo {author} {\bibfnamefont {B.}~\bibnamefont {Koczor}},
  \bibinfo {author} {\bibfnamefont {Z.}~\bibnamefont {Holmes}}, \bibinfo
  {author} {\bibfnamefont {E.-M.}\ \bibnamefont {Borrelli}}, \bibinfo {author}
  {\bibfnamefont {A.}~\bibnamefont {Gilyén}}, \bibinfo {author} {\bibfnamefont
  {H.-Y.}\ \bibnamefont {Huang}}, \bibinfo {author} {\bibfnamefont
  {Z.}~\bibnamefont {Cai}}, \bibinfo {author} {\bibfnamefont {A.}~\bibnamefont
  {Acín}}, \bibinfo {author} {\bibfnamefont {L.}~\bibnamefont {Aolita}},
  \bibinfo {author} {\bibfnamefont {L.}~\bibnamefont {Banchi}}, \bibinfo
  {author} {\bibfnamefont {F.~G. S.~L.}\ \bibnamefont {Brandão}}, \bibinfo
  {author} {\bibfnamefont {D.}~\bibnamefont {Cavalcanti}}, \bibinfo {author}
  {\bibfnamefont {T.}~\bibnamefont {Cubitt}}, \bibinfo {author} {\bibfnamefont
  {S.~N.}\ \bibnamefont {Filippov}}, \bibinfo {author} {\bibfnamefont
  {G.}~\bibnamefont {García-Pérez}}, \bibinfo {author} {\bibfnamefont
  {J.}~\bibnamefont {Goold}}, \bibinfo {author} {\bibfnamefont
  {O.}~\bibnamefont {Kálmán}}, \bibinfo {author} {\bibfnamefont
  {E.}~\bibnamefont {Kyoseva}}, \bibinfo {author} {\bibfnamefont {M.~A.~C.}\
  \bibnamefont {Rossi}}, \bibinfo {author} {\bibfnamefont {B.}~\bibnamefont
  {Sokolov}}, \bibinfo {author} {\bibfnamefont {I.}~\bibnamefont
  {Tavernelli}},\ and\ \bibinfo {author} {\bibfnamefont {S.}~\bibnamefont
  {Maniscalco}},\ }\href {https://arxiv.org/abs/2501.05694} {\bibinfo {title}
  {Myths around quantum computation before full fault tolerance: What no-go
  theorems rule out and what they don't}} (\bibinfo {year} {2025}),\ \Eprint
  {https://arxiv.org/abs/2501.05694} {arXiv:2501.05694} \BibitemShut {NoStop}%
\bibitem [{\citenamefont {Lu}\ \emph {et~al.}(2022)\citenamefont {Lu},
  \citenamefont {Zhou}, \citenamefont {Bao}, \citenamefont {Chen},
  \citenamefont {Li},\ and\ \citenamefont {Zhu}}]{lu2022dpm}%
  \BibitemOpen
  \bibfield  {author} {\bibinfo {author} {\bibfnamefont {C.}~\bibnamefont
  {Lu}}, \bibinfo {author} {\bibfnamefont {Y.}~\bibnamefont {Zhou}}, \bibinfo
  {author} {\bibfnamefont {F.}~\bibnamefont {Bao}}, \bibinfo {author}
  {\bibfnamefont {J.}~\bibnamefont {Chen}}, \bibinfo {author} {\bibfnamefont
  {C.}~\bibnamefont {Li}},\ and\ \bibinfo {author} {\bibfnamefont
  {J.}~\bibnamefont {Zhu}},\ }\bibfield  {title} {\bibinfo {title}
  {{DPM-solver: A fast ODE solver for diffusion probabilistic model sampling in
  around 10 steps}},\ }\href {https://doi.org/10.48550/arXiv.2206.00927}
  {\bibfield  {journal} {\bibinfo  {journal} {Adv. Neur. Inf. Proc. Sys.
  (NeurIPS)}\ }\textbf {\bibinfo {volume} {35}},\ \bibinfo {pages} {5775}
  (\bibinfo {year} {2022})}\BibitemShut {NoStop}%
\bibitem [{\citenamefont {Lu}\ \emph {et~al.}(2023)\citenamefont {Lu},
  \citenamefont {Zhou}, \citenamefont {Bao}, \citenamefont {Chen},
  \citenamefont {Li},\ and\ \citenamefont {Zhu}}]{lu2022dpm++}%
  \BibitemOpen
  \bibfield  {author} {\bibinfo {author} {\bibfnamefont {C.}~\bibnamefont
  {Lu}}, \bibinfo {author} {\bibfnamefont {Y.}~\bibnamefont {Zhou}}, \bibinfo
  {author} {\bibfnamefont {F.}~\bibnamefont {Bao}}, \bibinfo {author}
  {\bibfnamefont {J.}~\bibnamefont {Chen}}, \bibinfo {author} {\bibfnamefont
  {C.}~\bibnamefont {Li}},\ and\ \bibinfo {author} {\bibfnamefont
  {J.}~\bibnamefont {Zhu}},\ }\href {https://arxiv.org/abs/2211.01095}
  {\bibinfo {title} {{DPM-Solver++: Fast solver for guided sampling of
  diffusion probabilistic models}}} (\bibinfo {year} {2023}),\ \Eprint
  {https://arxiv.org/abs/2211.01095} {arXiv:2211.01095} \BibitemShut {NoStop}%
\bibitem [{\citenamefont {Zhao}\ \emph {et~al.}(2024)\citenamefont {Zhao},
  \citenamefont {Bai}, \citenamefont {Rao}, \citenamefont {Zhou},\ and\
  \citenamefont {Lu}}]{zhao2024unipc}%
  \BibitemOpen
  \bibfield  {author} {\bibinfo {author} {\bibfnamefont {W.}~\bibnamefont
  {Zhao}}, \bibinfo {author} {\bibfnamefont {L.}~\bibnamefont {Bai}}, \bibinfo
  {author} {\bibfnamefont {Y.}~\bibnamefont {Rao}}, \bibinfo {author}
  {\bibfnamefont {J.}~\bibnamefont {Zhou}},\ and\ \bibinfo {author}
  {\bibfnamefont {J.}~\bibnamefont {Lu}},\ }\bibfield  {title} {\bibinfo
  {title} {{UniPC: A unified predictor-corrector framework for fast sampling of
  diffusion models}},\ }\href {https://doi.org/10.48550/arXiv.2302.04867}
  {\bibfield  {journal} {\bibinfo  {journal} {Adv. Neur. Inf. Proc. Sys.
  (NeurIPS)}\ }\textbf {\bibinfo {volume} {36}},\ \bibinfo {pages} {49842}
  (\bibinfo {year} {2024})}\BibitemShut {NoStop}%
\bibitem [{\citenamefont {Liu}\ \emph {et~al.}(2024)\citenamefont {Liu},
  \citenamefont {Liu}, \citenamefont {Liu}, \citenamefont {Ye}, \citenamefont
  {Wang}, \citenamefont {Alexeev}, \citenamefont {Eisert},\ and\ \citenamefont
  {Jiang}}]{Liu:2023coc}%
  \BibitemOpen
  \bibfield  {author} {\bibinfo {author} {\bibfnamefont {J.}~\bibnamefont
  {Liu}}, \bibinfo {author} {\bibfnamefont {M.}~\bibnamefont {Liu}}, \bibinfo
  {author} {\bibfnamefont {J.-P.}\ \bibnamefont {Liu}}, \bibinfo {author}
  {\bibfnamefont {Z.}~\bibnamefont {Ye}}, \bibinfo {author} {\bibfnamefont
  {Y.}~\bibnamefont {Wang}}, \bibinfo {author} {\bibfnamefont {Y.}~\bibnamefont
  {Alexeev}}, \bibinfo {author} {\bibfnamefont {J.}~\bibnamefont {Eisert}},\
  and\ \bibinfo {author} {\bibfnamefont {L.}~\bibnamefont {Jiang}},\ }\bibfield
   {title} {\bibinfo {title} {{Towards provably efficient quantum algorithms
  for large-scale machine-learning models}},\ }\href
  {https://doi.org/10.1038/s41467-023-43957-x} {\bibfield  {journal} {\bibinfo
  {journal} {Nature Comm.}\ }\textbf {\bibinfo {volume} {15}},\ \bibinfo
  {pages} {434} (\bibinfo {year} {2024})},\ \Eprint
  {https://arxiv.org/abs/2303.03428} {arXiv:2303.03428} \BibitemShut {NoStop}%
\bibitem [{\citenamefont {Liu}\ \emph {et~al.}(2021{\natexlab{b}})\citenamefont
  {Liu}, \citenamefont {Kolden}, \citenamefont {Krovi}, \citenamefont
  {Loureiro}, \citenamefont {Trivisa},\ and\ \citenamefont
  {Childs}}]{liu2021efficient}%
  \BibitemOpen
  \bibfield  {author} {\bibinfo {author} {\bibfnamefont {J.-P.}\ \bibnamefont
  {Liu}}, \bibinfo {author} {\bibfnamefont {H.~{\O}.}\ \bibnamefont {Kolden}},
  \bibinfo {author} {\bibfnamefont {H.~K.}\ \bibnamefont {Krovi}}, \bibinfo
  {author} {\bibfnamefont {N.~F.}\ \bibnamefont {Loureiro}}, \bibinfo {author}
  {\bibfnamefont {K.}~\bibnamefont {Trivisa}},\ and\ \bibinfo {author}
  {\bibfnamefont {A.~M.}\ \bibnamefont {Childs}},\ }\bibfield  {title}
  {\bibinfo {title} {Efficient quantum algorithm for dissipative nonlinear
  differential equations},\ }\href {https://doi.org/10.1073/pnas.2026805118}
  {\bibfield  {journal} {\bibinfo  {journal} {Proc. Natl. Ac. Sc.}\ }\textbf
  {\bibinfo {volume} {118}},\ \bibinfo {pages} {e2026805118} (\bibinfo {year}
  {2021}{\natexlab{b}})},\ \bibinfo {note}
  {\href{https://arxiv.org/abs/2011.03185}{arXiv:2011.03185}}\BibitemShut
  {NoStop}%
\bibitem [{\citenamefont {Ambainis}(2010)}]{ambainis2010variable}%
  \BibitemOpen
  \bibfield  {author} {\bibinfo {author} {\bibfnamefont {A.}~\bibnamefont
  {Ambainis}},\ }\href {https://arxiv.org/abs/1010.4458} {\bibinfo {title}
  {Variable time amplitude amplification and a faster quantum algorithm for
  solving systems of linear equations}} (\bibinfo {year} {2010}),\ \Eprint
  {https://arxiv.org/abs/1010.4458} {arXiv:1010.4458} \BibitemShut {NoStop}%
\bibitem [{\citenamefont {Childs}\ \emph {et~al.}(2017)\citenamefont {Childs},
  \citenamefont {Kothari},\ and\ \citenamefont {Somma}}]{Childs_2017}%
  \BibitemOpen
  \bibfield  {author} {\bibinfo {author} {\bibfnamefont {A.~M.}\ \bibnamefont
  {Childs}}, \bibinfo {author} {\bibfnamefont {R.}~\bibnamefont {Kothari}},\
  and\ \bibinfo {author} {\bibfnamefont {R.~D.}\ \bibnamefont {Somma}},\
  }\bibfield  {title} {\bibinfo {title} {Quantum algorithm for systems of
  linear equations with exponentially improved dependence on precision},\
  }\href {https://doi.org/10.1137/16m1087072} {\bibfield  {journal} {\bibinfo
  {journal} {SIAM J. Comp.}\ }\textbf {\bibinfo {volume} {46}},\ \bibinfo
  {pages} {1920–1950} (\bibinfo {year} {2017})}\BibitemShut {NoStop}%
\bibitem [{\citenamefont {Low}\ and\ \citenamefont
  {Su}(2024)}]{low2024QLAOptimal}%
  \BibitemOpen
  \bibfield  {author} {\bibinfo {author} {\bibfnamefont {G.~H.}\ \bibnamefont
  {Low}}\ and\ \bibinfo {author} {\bibfnamefont {Y.}~\bibnamefont {Su}},\
  }\href {https://arxiv.org/abs/2410.18178} {\bibinfo {title} {Quantum linear
  system algorithm with optimal queries to initial state preparation}}
  (\bibinfo {year} {2024}),\ \Eprint {https://arxiv.org/abs/2410.18178}
  {arXiv:2410.18178} \BibitemShut {NoStop}%
\bibitem [{\citenamefont {Costa}\ \emph {et~al.}(2021)\citenamefont {Costa},
  \citenamefont {An}, \citenamefont {Sanders}, \citenamefont {Su},
  \citenamefont {Babbush},\ and\ \citenamefont {Berry}}]{costa2021optimal}%
  \BibitemOpen
  \bibfield  {author} {\bibinfo {author} {\bibfnamefont {P.~C.~S.}\
  \bibnamefont {Costa}}, \bibinfo {author} {\bibfnamefont {D.}~\bibnamefont
  {An}}, \bibinfo {author} {\bibfnamefont {Y.~R.}\ \bibnamefont {Sanders}},
  \bibinfo {author} {\bibfnamefont {Y.}~\bibnamefont {Su}}, \bibinfo {author}
  {\bibfnamefont {R.}~\bibnamefont {Babbush}},\ and\ \bibinfo {author}
  {\bibfnamefont {D.~W.}\ \bibnamefont {Berry}},\ }\href
  {https://arxiv.org/abs/2111.08152} {\bibinfo {title} {Optimal scaling quantum
  linear systems solver via discrete adiabatic theorem}} (\bibinfo {year}
  {2021}),\ \Eprint {https://arxiv.org/abs/2111.08152} {arXiv:2111.08152}
  \BibitemShut {NoStop}%
\bibitem [{\citenamefont {Dalzell}(2024)}]{dalzell2024shortcut}%
  \BibitemOpen
  \bibfield  {author} {\bibinfo {author} {\bibfnamefont {A.~M.}\ \bibnamefont
  {Dalzell}},\ }\href {https://arxiv.org/abs/2406.12086} {\bibinfo {title} {A
  shortcut to an optimal quantum linear system solver}} (\bibinfo {year}
  {2024}),\ \Eprint {https://arxiv.org/abs/2406.12086} {arXiv:2406.12086}
  \BibitemShut {NoStop}%
\bibitem [{\citenamefont {Gilyén}\ \emph {et~al.}(2019)\citenamefont
  {Gilyén}, \citenamefont {Su}, \citenamefont {Low},\ and\ \citenamefont
  {Wiebe}}]{Gily_n_2019}%
  \BibitemOpen
  \bibfield  {author} {\bibinfo {author} {\bibfnamefont {A.}~\bibnamefont
  {Gilyén}}, \bibinfo {author} {\bibfnamefont {Y.}~\bibnamefont {Su}},
  \bibinfo {author} {\bibfnamefont {G.~H.}\ \bibnamefont {Low}},\ and\ \bibinfo
  {author} {\bibfnamefont {N.}~\bibnamefont {Wiebe}},\ }\bibfield  {title}
  {\bibinfo {title} {Quantum singular value transformation and beyond:
  exponential improvements for quantum matrix arithmetics},\ }in\ \href
  {https://doi.org/10.1145/3313276.3316366} {\emph {\bibinfo {booktitle}
  {Proceedings of the 51st Annual ACM SIGACT Symposium on Theory of
  Computing}}},\ \bibinfo {series and number} {STOC ’19}\ (\bibinfo
  {publisher} {ACM},\ \bibinfo {year} {2019})\BibitemShut {NoStop}%
\bibitem [{\citenamefont {An}\ \emph {et~al.}(2023{\natexlab{a}})\citenamefont
  {An}, \citenamefont {Liu},\ and\ \citenamefont {Lin}}]{An_2023}%
  \BibitemOpen
  \bibfield  {author} {\bibinfo {author} {\bibfnamefont {D.}~\bibnamefont
  {An}}, \bibinfo {author} {\bibfnamefont {J.-P.}\ \bibnamefont {Liu}},\ and\
  \bibinfo {author} {\bibfnamefont {L.}~\bibnamefont {Lin}},\ }\bibfield
  {title} {\bibinfo {title} {{Linear combination of Hamiltonian simulation for
  nonunitary dynamics with optimal state preparation cost}},\ }\href
  {https://doi.org/10.1103/physrevlett.131.150603} {\bibfield  {journal}
  {\bibinfo  {journal} {Phys. Rev. Lett.}\ }\textbf {\bibinfo {volume} {131}},\
  \bibinfo {pages} {150603} (\bibinfo {year} {2023}{\natexlab{a}})}\BibitemShut
  {NoStop}%
\bibitem [{\citenamefont {An}\ \emph {et~al.}(2023{\natexlab{b}})\citenamefont
  {An}, \citenamefont {Childs},\ and\ \citenamefont {Lin}}]{an2023quantum}%
  \BibitemOpen
  \bibfield  {author} {\bibinfo {author} {\bibfnamefont {D.}~\bibnamefont
  {An}}, \bibinfo {author} {\bibfnamefont {A.~M.}\ \bibnamefont {Childs}},\
  and\ \bibinfo {author} {\bibfnamefont {L.}~\bibnamefont {Lin}},\ }\href
  {https://arxiv.org/abs/2312.03916} {\bibinfo {title} {Quantum algorithm for
  linear non-unitary dynamics with near-optimal dependence on all parameters}}
  (\bibinfo {year} {2023}{\natexlab{b}}),\ \Eprint
  {https://arxiv.org/abs/2312.03916} {arXiv:2312.03916} \BibitemShut {NoStop}%
\bibitem [{\citenamefont {Childs}\ and\ \citenamefont
  {Wiebe}(2012)}]{Childs2012Hamiltonian}%
  \BibitemOpen
  \bibfield  {author} {\bibinfo {author} {\bibfnamefont {A.~M.}\ \bibnamefont
  {Childs}}\ and\ \bibinfo {author} {\bibfnamefont {N.}~\bibnamefont {Wiebe}},\
  }\bibfield  {title} {\bibinfo {title} {Hamiltonian simulation using linear
  combinations of unitary operations},\ }\href
  {https://doi.org/10.26421/qic12.11-12} {\bibfield  {journal} {\bibinfo
  {journal} {Quant. Inf. Comp.}\ }\textbf {\bibinfo {volume} {12}},\ \bibinfo
  {pages} {0901} (\bibinfo {year} {2012})}\BibitemShut {NoStop}%
\bibitem [{\citenamefont {Hann}\ \emph {et~al.}(2019)\citenamefont {Hann},
  \citenamefont {Zou}, \citenamefont {Zhang}, \citenamefont {Chu},
  \citenamefont {Schoelkopf}, \citenamefont {Girvin},\ and\ \citenamefont
  {Jiang}}]{hann2019hardware}%
  \BibitemOpen
  \bibfield  {author} {\bibinfo {author} {\bibfnamefont {C.~T.}\ \bibnamefont
  {Hann}}, \bibinfo {author} {\bibfnamefont {C.-L.}\ \bibnamefont {Zou}},
  \bibinfo {author} {\bibfnamefont {Y.}~\bibnamefont {Zhang}}, \bibinfo
  {author} {\bibfnamefont {Y.}~\bibnamefont {Chu}}, \bibinfo {author}
  {\bibfnamefont {R.~J.}\ \bibnamefont {Schoelkopf}}, \bibinfo {author}
  {\bibfnamefont {S.~M.}\ \bibnamefont {Girvin}},\ and\ \bibinfo {author}
  {\bibfnamefont {L.}~\bibnamefont {Jiang}},\ }\bibfield  {title} {\bibinfo
  {title} {Hardware-efficient quantum random access memory with hybrid quantum
  acoustic systems},\ }\href {https://doi.org/10.1103/PhysRevLett.123.250501}
  {\bibfield  {journal} {\bibinfo  {journal} {Phys. Rev. Lett.}\ }\textbf
  {\bibinfo {volume} {123}},\ \bibinfo {pages} {250501} (\bibinfo {year}
  {2019})}\BibitemShut {NoStop}%
\bibitem [{\citenamefont {Matteo}\ \emph {et~al.}(2020)\citenamefont {Matteo},
  \citenamefont {Gheorghiu},\ and\ \citenamefont {Mosca}}]{Matteo_2020}%
  \BibitemOpen
  \bibfield  {author} {\bibinfo {author} {\bibfnamefont {O.~D.}\ \bibnamefont
  {Matteo}}, \bibinfo {author} {\bibfnamefont {V.}~\bibnamefont {Gheorghiu}},\
  and\ \bibinfo {author} {\bibfnamefont {M.}~\bibnamefont {Mosca}},\ }\bibfield
   {title} {\bibinfo {title} {Fault-tolerant resource estimation of quantum
  random-access memories},\ }\href {https://doi.org/10.1109/tqe.2020.2965803}
  {\bibfield  {journal} {\bibinfo  {journal} {IEEE Trans. Quant. Eng.}\
  }\textbf {\bibinfo {volume} {1}},\ \bibinfo {pages} {1–13} (\bibinfo {year}
  {2020})}\BibitemShut {NoStop}%
\bibitem [{\citenamefont {Hann}\ \emph {et~al.}(2021)\citenamefont {Hann},
  \citenamefont {Lee}, \citenamefont {Girvin},\ and\ \citenamefont
  {Jiang}}]{Hann_2021}%
  \BibitemOpen
  \bibfield  {author} {\bibinfo {author} {\bibfnamefont {C.~T.}\ \bibnamefont
  {Hann}}, \bibinfo {author} {\bibfnamefont {G.}~\bibnamefont {Lee}}, \bibinfo
  {author} {\bibfnamefont {S.}~\bibnamefont {Girvin}},\ and\ \bibinfo {author}
  {\bibfnamefont {L.}~\bibnamefont {Jiang}},\ }\bibfield  {title} {\bibinfo
  {title} {Resilience of quantum random access memory to generic noise},\
  }\href {https://doi.org/10.1103/prxquantum.2.020311} {\bibfield  {journal}
  {\bibinfo  {journal} {PRX Quantum}\ }\textbf {\bibinfo {volume} {2}},\
  \bibinfo {pages} {020311} (\bibinfo {year} {2021})}\BibitemShut {NoStop}%
\bibitem [{\citenamefont {Wang}\ \emph {et~al.}(2024)\citenamefont {Wang},
  \citenamefont {Alexeev}, \citenamefont {Jiang}, \citenamefont {Chong},\ and\
  \citenamefont {Liu}}]{Wang:2023oon}%
  \BibitemOpen
  \bibfield  {author} {\bibinfo {author} {\bibfnamefont {Y.}~\bibnamefont
  {Wang}}, \bibinfo {author} {\bibfnamefont {Y.}~\bibnamefont {Alexeev}},
  \bibinfo {author} {\bibfnamefont {L.}~\bibnamefont {Jiang}}, \bibinfo
  {author} {\bibfnamefont {F.~T.}\ \bibnamefont {Chong}},\ and\ \bibinfo
  {author} {\bibfnamefont {J.}~\bibnamefont {Liu}},\ }\bibfield  {title}
  {\bibinfo {title} {{Fundamental causal bounds of quantum random access
  memories}},\ }\href {https://doi.org/10.1038/s41534-024-00848-3} {\bibfield
  {journal} {\bibinfo  {journal} {npj Quant. Inf.}\ }\textbf {\bibinfo {volume}
  {10}},\ \bibinfo {pages} {71} (\bibinfo {year} {2024})},\ \Eprint
  {https://arxiv.org/abs/2307.13460} {arXiv:2307.13460 [quant-ph]} \BibitemShut
  {NoStop}%
\bibitem [{\citenamefont {Eisert}\ \emph {et~al.}(2020)\citenamefont {Eisert},
  \citenamefont {Hangleiter}, \citenamefont {Walk}, \citenamefont {Roth},
  \citenamefont {Markham}, \citenamefont {Parekh}, \citenamefont {Chabaud},\
  and\ \citenamefont {Kashefi}}]{BenchmarkingReview}%
  \BibitemOpen
  \bibfield  {author} {\bibinfo {author} {\bibfnamefont {J.}~\bibnamefont
  {Eisert}}, \bibinfo {author} {\bibfnamefont {D.}~\bibnamefont {Hangleiter}},
  \bibinfo {author} {\bibfnamefont {N.}~\bibnamefont {Walk}}, \bibinfo {author}
  {\bibfnamefont {I.}~\bibnamefont {Roth}}, \bibinfo {author} {\bibfnamefont
  {D.}~\bibnamefont {Markham}}, \bibinfo {author} {\bibfnamefont
  {R.}~\bibnamefont {Parekh}}, \bibinfo {author} {\bibfnamefont
  {U.}~\bibnamefont {Chabaud}},\ and\ \bibinfo {author} {\bibfnamefont
  {E.}~\bibnamefont {Kashefi}},\ }\bibfield  {title} {\bibinfo {title} {Quantum
  certification and benchmarking},\ }\href
  {https://doi.org/10.1038/s42254-020-0186-4} {\bibfield  {journal} {\bibinfo
  {journal} {Nature Rev. Phys.}\ }\textbf {\bibinfo {volume} {2}},\ \bibinfo
  {pages} {382} (\bibinfo {year} {2020})}\BibitemShut {NoStop}%
\bibitem [{\citenamefont {Aaronson}(2018)}]{aaronson2018shadowtomography}%
  \BibitemOpen
  \bibfield  {author} {\bibinfo {author} {\bibfnamefont {S.}~\bibnamefont
  {Aaronson}},\ }\href {https://arxiv.org/abs/1711.01053} {\bibinfo {title}
  {Shadow tomography of quantum states}} (\bibinfo {year} {2018}),\ \Eprint
  {https://arxiv.org/abs/1711.01053} {arXiv:1711.01053} \BibitemShut {NoStop}%
\bibitem [{\citenamefont {Huang}\ \emph {et~al.}(2020)\citenamefont {Huang},
  \citenamefont {Kueng},\ and\ \citenamefont {Preskill}}]{Huang_2020}%
  \BibitemOpen
  \bibfield  {author} {\bibinfo {author} {\bibfnamefont {H.-Y.}\ \bibnamefont
  {Huang}}, \bibinfo {author} {\bibfnamefont {R.}~\bibnamefont {Kueng}},\ and\
  \bibinfo {author} {\bibfnamefont {J.}~\bibnamefont {Preskill}},\ }\bibfield
  {title} {\bibinfo {title} {Predicting many properties of a quantum system
  from very few measurements},\ }\href
  {https://doi.org/10.1038/s41567-020-0932-7} {\bibfield  {journal} {\bibinfo
  {journal} {Nature Phys.}\ }\textbf {\bibinfo {volume} {16}},\ \bibinfo
  {pages} {1050–1057} (\bibinfo {year} {2020})}\BibitemShut {NoStop}%
\bibitem [{\citenamefont {Bertoni}\ \emph {et~al.}(2024)\citenamefont
  {Bertoni}, \citenamefont {Haferkamp}, \citenamefont {Hinsche}, \citenamefont
  {Ioannou}, \citenamefont {Eisert},\ and\ \citenamefont
  {Pashayan}}]{ShallowShadows}%
  \BibitemOpen
  \bibfield  {author} {\bibinfo {author} {\bibfnamefont {C.}~\bibnamefont
  {Bertoni}}, \bibinfo {author} {\bibfnamefont {J.}~\bibnamefont {Haferkamp}},
  \bibinfo {author} {\bibfnamefont {M.}~\bibnamefont {Hinsche}}, \bibinfo
  {author} {\bibfnamefont {M.}~\bibnamefont {Ioannou}}, \bibinfo {author}
  {\bibfnamefont {J.}~\bibnamefont {Eisert}},\ and\ \bibinfo {author}
  {\bibfnamefont {H.}~\bibnamefont {Pashayan}},\ }\bibfield  {title} {\bibinfo
  {title} {{Shallow shadows: Expectation estimation using low-depth random
  Clifford circuits}},\ }\href {https://doi.org/10.1103/PhysRevLett.133.020602}
  {\bibfield  {journal} {\bibinfo  {journal} {Phys. Rev. Lett.}\ }\textbf
  {\bibinfo {volume} {133}},\ \bibinfo {pages} {020602} (\bibinfo {year}
  {2024})}\BibitemShut {NoStop}%
\bibitem [{\citenamefont {Bao}\ \emph {et~al.}(2023)\citenamefont {Bao},
  \citenamefont {Nie}, \citenamefont {Xue}, \citenamefont {Cao}, \citenamefont
  {Li}, \citenamefont {Su},\ and\ \citenamefont {Zhu}}]{bao2022all}%
  \BibitemOpen
  \bibfield  {author} {\bibinfo {author} {\bibfnamefont {F.}~\bibnamefont
  {Bao}}, \bibinfo {author} {\bibfnamefont {S.}~\bibnamefont {Nie}}, \bibinfo
  {author} {\bibfnamefont {K.}~\bibnamefont {Xue}}, \bibinfo {author}
  {\bibfnamefont {Y.}~\bibnamefont {Cao}}, \bibinfo {author} {\bibfnamefont
  {C.}~\bibnamefont {Li}}, \bibinfo {author} {\bibfnamefont {H.}~\bibnamefont
  {Su}},\ and\ \bibinfo {author} {\bibfnamefont {J.}~\bibnamefont {Zhu}},\
  }\bibfield  {title} {\bibinfo {title} {All are worth words: A vit backbone
  for diffusion models},\ }in\ \href@noop {} {\emph {\bibinfo {booktitle}
  {CVPR}}}\ (\bibinfo {year} {2023})\BibitemShut {NoStop}%
\bibitem [{\citenamefont {Rombach}\ \emph {et~al.}(2021)\citenamefont
  {Rombach}, \citenamefont {Blattmann}, \citenamefont {Lorenz}, \citenamefont
  {Esser},\ and\ \citenamefont {Ommer}}]{rombach2021highresolution}%
  \BibitemOpen
  \bibfield  {author} {\bibinfo {author} {\bibfnamefont {R.}~\bibnamefont
  {Rombach}}, \bibinfo {author} {\bibfnamefont {A.}~\bibnamefont {Blattmann}},
  \bibinfo {author} {\bibfnamefont {D.}~\bibnamefont {Lorenz}}, \bibinfo
  {author} {\bibfnamefont {P.}~\bibnamefont {Esser}},\ and\ \bibinfo {author}
  {\bibfnamefont {B.}~\bibnamefont {Ommer}},\ }\href@noop {} {\bibinfo {title}
  {High-resolution image synthesis with latent diffusion models}} (\bibinfo
  {year} {2021}),\ \Eprint {https://arxiv.org/abs/2112.10752}
  {arXiv:2112.10752} \BibitemShut {NoStop}%
\bibitem [{\citenamefont {Deng}\ \emph {et~al.}(2009)\citenamefont {Deng},
  \citenamefont {Dong}, \citenamefont {Socher}, \citenamefont {Li},
  \citenamefont {Li},\ and\ \citenamefont {Fei-Fei}}]{deng2009imagenet}%
  \BibitemOpen
  \bibfield  {author} {\bibinfo {author} {\bibfnamefont {J.}~\bibnamefont
  {Deng}}, \bibinfo {author} {\bibfnamefont {W.}~\bibnamefont {Dong}}, \bibinfo
  {author} {\bibfnamefont {R.}~\bibnamefont {Socher}}, \bibinfo {author}
  {\bibfnamefont {L.-J.}\ \bibnamefont {Li}}, \bibinfo {author} {\bibfnamefont
  {K.}~\bibnamefont {Li}},\ and\ \bibinfo {author} {\bibfnamefont
  {L.}~\bibnamefont {Fei-Fei}},\ }\bibfield  {title} {\bibinfo {title}
  {Imagenet: A large-scale hierarchical image database},\ }in\ \href@noop {}
  {\emph {\bibinfo {booktitle} {2009 IEEE conference on computer vision and
  pattern recognition}}}\ (\bibinfo {organization} {Ieee},\ \bibinfo {year}
  {2009})\ pp.\ \bibinfo {pages} {248--255}\BibitemShut {NoStop}%
\end{thebibliography}

%

\end{document}